\documentclass[%
 aip,
 amsmath,amssymb,
 reprint,%
]{revtex4-1}

\usepackage{graphicx}
\usepackage{dcolumn}
\usepackage{bm}

\usepackage[utf8]{inputenc}
\usepackage[T1]{fontenc}
\usepackage[english]{babel}
\usepackage{mathptmx}
\usepackage{etoolbox}

\usepackage{epstopdf}
\usepackage{siunitx}
\usepackage{booktabs}
\usepackage[normalem]{ulem} 
\usepackage{threeparttable} 
\usepackage{tabularx}
\usepackage{adjustbox}
\usepackage{verbatim}
\usepackage[colorinlistoftodos]{todonotes}
\usepackage{pbox}
\usepackage[figurename=Fig.,tablename=Tab., justification=raggedright, format=plain]{caption}
\usepackage[caption=false]{subfig}
\usepackage{hyperref}
\usepackage[capitalise]{cleveref}
\usepackage{parskip}
\usepackage{rotating}

\newcommand{\dquotes}[1]{``#1''} 

\hypersetup{
    colorlinks=true,
    linkcolor=blue,
    citecolor=blue,
    filecolor=blue,
    urlcolor=blue,
}

\crefformat{equation}{Eq.~(#2#1#3)} 
\makeatletter
\def\@email#1#2{%
 \endgroup
 \patchcmd{\titleblock@produce}
  {\frontmatter@RRAPformat}
  {\frontmatter@RRAPformat{\produce@RRAP{*#1\href{mailto:#2}{#2}}}\frontmatter@RRAPformat}
  {}{}
}%
\makeatother
\begin{document}

\preprint{AIP/123-QED}

\title{Study of dielectric breakdown in liquid xenon with the XeBrA experiment}

\author{J. Watson}
  \affiliation{University of California, Berkeley, Berkeley, California 94720, USA} 
  \affiliation{Lawrence Berkeley National Laboratory, Berkeley, California 94720, USA}
  \email{watson\_james@berkeley.edu}
\author{I. Olcina}
  \affiliation{University of California, Berkeley, Berkeley, California 94720, USA} 
  \affiliation{Lawrence Berkeley National Laboratory, Berkeley, California 94720, USA}
  \email[Author to whom correspondence should be addressed: ]{ibles10@berkeley.edu}
\author{J. Soria}
  \affiliation{University of California, Berkeley, Berkeley, California 94720, USA} 
  \affiliation{Lawrence Berkeley National Laboratory, Berkeley, California 94720, USA}
\author{D.N.~McKinsey}
  \affiliation{University of California, Berkeley, Berkeley, California 94720, USA} 
  \affiliation{Lawrence Berkeley National Laboratory, Berkeley, California 94720, USA}
\author{S.~Kravitz} 
  \affiliation{Lawrence Berkeley National Laboratory, Berkeley, California 94720, USA}
\author{E. E.~Deck}
  \altaffiliation[Now at: ]{Los Alamos National Laboratory, Los Alamos, New Mexico 87545, USA}
  \affiliation{University of California, Berkeley, Berkeley, California 94720, USA} 
  \affiliation{Lawrence Berkeley National Laboratory, Berkeley, California 94720, USA}
\author{E. P.~Bernard}
  \altaffiliation[Now at: ]{Lawrence Livermore National Laboratory, Livermore, California 94550, USA}
  \affiliation{University of California, Berkeley, Berkeley, California 94720, USA} 
  \affiliation{Lawrence Berkeley National Laboratory, Berkeley, California 94720, USA}
\author{L.~Tvrznikova}
  \altaffiliation[Now at: ]{Waymo LLC, Mountain View, CA 94043, USA}
  \affiliation{University of California, Berkeley, Berkeley, California 94720, USA} 
  \affiliation{Lawrence Berkeley National Laboratory, Berkeley, California 94720, USA}
  \affiliation{Yale University, New Haven, Connecticut 06511, USA} 
  \affiliation{Lawrence Livermore National Laboratory, Livermore, California 94550, USA}
\author{W.L.~Waldron}
  \affiliation{Lawrence Berkeley National Laboratory, Berkeley, California 94720, USA}
\author{Q.~Riffard}
  \altaffiliation[Now at: ]{Nuro, 1300 Terra Bella Ave 200, Mountain View, California 94043, USA}
  \affiliation{Lawrence Berkeley National Laboratory, Berkeley, California 94720, USA}
\author{K.O'Sullivan}
  \altaffiliation[Now at: ]{Grammarly Inc., San Francisco, CA 94104, USA}
  \affiliation{University of California, Berkeley, Berkeley, California 94720, USA} 
  \affiliation{Lawrence Berkeley National Laboratory, Berkeley, California 94720, USA}

\date{\today}

\begin{abstract}
Maintaining the electric fields necessary for the current generation of noble liquid time projection chambers (TPCs), with drift lengths exceeding one meter, requires a large negative voltage applied to their cathode. Delivering such high voltage is associated with an elevated risk of electrostatic discharge and electroluminescence, which would be detrimental to the performance of the experiment.
The Xenon Breakdown Apparatus (XeBrA) is a five-liter, high voltage test chamber built to investigate the contributing factors to electrical breakdown in noble liquids. In this work, we present the main findings after conducting scans over stressed electrode areas, surface finish, pressure, and high voltage ramp speed in the medium of liquid xenon.
Area scaling and surface finish were observed to be the dominant factors affecting breakdown, whereas no significant changes were observed with varying pressure or ramp speed.
A general rise in both anode current and photon rate was observed in the last 30 seconds leading up to a breakdown, with a marked increase in the last couple of seconds. In addition, the position of breakdowns was reconstructed with a system of high-speed cameras and a moderate correlation with the Fowler-Nordheim field emission model was found. The system of cameras also allowed us to find tentative evidence for bubble nucleation being the originating mechanism of breakdown in the liquid. We deem the results presented in this work to be of particular interest for the design of future, large TPCs, and practical recommendations are provided.
\end{abstract}

\maketitle

%

\section{Introduction}
\label{sec:intro} 
Dielectric breakdown in time projection chambers (TPCs) is a cause of concern because it could damage the detector and affect its overall performance. Therefore, the characterization of dielectric breakdown in noble liquids, especially in liquid argon (LAr) and liquid xenon (LXe), is of great importance not only for the current generation of experiments but also for the design of next-generation experiments such as DarkSide-20k~\cite{Aalseth:2017fik}, DUNE~\cite{Abi:2018dnh}, DARWIN~\cite{Aalbers:2016jon}, and nEXO~\cite{Albert:2017hjq}. Recently, PandaX-II observed elevated photon rates~\cite{PandaX-II:2022waa}, highlighting the need to understand these phenomena.

In LXe TPCs, the drift fields are typically tens to hundreds of V/cm, and an order of magnitude higher surrounding the high voltage (HV) feedthroughs and in the extraction region in dual-phase TPCs.  
Because of this, the dielectric strength of LXe must be well understood so that control measures, either in the design of an experiment or during its operation, can be used to mitigate the risk of dielectric breakdown.
This is the goal of the Xenon Breakdown Apparatus (XeBrA), located at the Lawrence Berkeley National Laboratory (LBNL).
In this work, we report on measurements of voltage breakdown in LXe and provide an update on the work that was started in Ref.~\cite{tvrznikova_direct_2019}. 

The paper is organized as follows: \cref{sec:dependences} outlines the main factors affecting dielectric breakdown in LXe and \cref{sec:hazard_functions} presents the main semi-empirical models that were explored to test these dependencies.
\cref{sec:field_emission} introduces the Fowler-Nordheim (FN) model, used to parameterize field emission current in our system, and \cref{sec:xebra} gives a brief description of the XeBrA experiment. 
Both the experimental methods and simulation analysis are described in \cref{sec:methods} and the body of results is presented in \cref{sec:Results}. The conclusions are presented in \cref{sec:Conclusions}.

\subsection{Factors affecting breakdown}
\label{sec:dependences}
A \emph{breakdown} is an instability which occurs when an insulating medium becomes conducting at a sufficiently high voltage.
This can happen when an electron drifting through the medium, in addition to freeing other electrons through primary ionization, generates one or more net electrons on average through secondary ionization (e.g.~via photoionization or positive ions impinging on the cathodic surface). 
This leads to an exponential rise in current over time that, if unstopped, culminates into a breakdown.
When there is significant electron multiplication, the space charge from the electron and ion clouds generates an additional field.
If this space charge-induced field becomes of the same order as the bulk field, a significant rise in current occurs over a single drift length, a phenomenon known as \emph{streamer discharge}~\cite{streamer}.
%
%
In liquid xenon, streamer discharge is believed to be the primary breakdown mechanism due to its extremely slow ionic drift speed~\cite{hilt_ionic_1994}.

Based on simple density scaling of gas phase data, xenon is predicted to have a large bulk breakdown field of approximately \SI{1}{MV/cm}~\cite{Massarczyk_2017,Derenzo:1974zz} in liquid.
However, LXe has been empirically observed to break down at fields much lower than this (hundreds of \SI{}{kV/cm}~\cite{Rebel:2014uia,tvrznikova_direct_2019}).
Therefore, other mechanisms must contribute to the positive feedback necessary to achieve dielectric breakdown. 
Due to the high dielectric strength in the liquid, surface effects become the most likely explanation.
For instance, the presence of asperities on the electrodes could enhance the local electric field and increase electron multiplication to the point of streamer onset~\cite{Phan:2020nhz}.


The surface area of the electrodes exposed to high fields is referred to as the \textit{stressed electrode area} (SEA).
As SEA increases, the number of asperities that could initiate a breakdown increases as well.
In fact, an inversely proportional dependence of breakdown field with stressed area has been observed in several media~\cite{Auger:2015xlo,KAWASHIMA1974217,weber1956nitrogen,goshime1995weibull,GERHOLD1994579}.
Other dependencies have been noted as well, the main ones being:

\begin{enumerate}
    \item \textbf{Volume:} if the medium itself is a point of failure, then the breakdown voltage would decrease in proportion to volume.
    This can occur in cases in which high fields cause chemical reactions to occur or ionic bonds to dissociate. In noble liquids, neither are factors. However, one might observe such an effect in cases where bubbles are present throughout the stressed volume (if the medium is very close to the boiling curve).
    \item \textbf{Purity:} electrons drifting in liquid attach to electronegative impurities, such as oxygen.
    The reduction in charge decreases the  likelihood of triggering a breakdown. 
    \item \textbf{Pressure:} bubble nucleation and growth are suppressed by increasing the liquid pressure.
    It has been observed that increasing the liquid pressure resulted in higher breakdown voltages in both liquid nitrogen (LN) \cite{hayakawa_1997} and liquid helium (LHe) \cite{Gerhold_LiquidHelium,Phan:2020nhz}.
    \item \textbf{Ramp speed:} if the risk of breakdown is represented as a differential breakdown rate per unit time, then different ramp speeds might result in different  maximum breakdown voltages. 
    This has been observed in borosilicate glass and alumina~\cite{fischer_influence_2021}, ferrofluid~\cite{ ferrofluid}, and oil-impregnated pressboard\cite{Mohammed2013WeibullSI}.
    However, an opposite correlation was found in thin film polymers~\cite{thinfilm_ramprate}.
    \item \textbf{Surface finish:} higher sustained electric fields have been observed in more finely polished electrodes in both LHe~\cite{GERHOLD1994579,Gerhold1989,Phan:2020nhz} and LN$_2$~\cite{goshima1995nitrogen}.
    Furthermore, acid passivation and electropolishing have been shown to be effective methods at reducing emission rates in stainless steel wires~\cite{TOMAS201849}.
\end{enumerate}

\subsection{Hazard functions}
\label{sec:hazard_functions}

In this section, we present a brief overview of the reliability analysis that was performed in this work. The \textit{hazard function} $h(t)$ is defined as the differential failure rate at a given point in time $t$, conditional on surviving up to that point\cite{miller_survival_2011}.
Hence, the \textit{survival function} $S(t)$, defined as the probability of ramping to time $t$ without a breakdown, can be represented as
\begin{equation}
    S(t) = \exp\left(-\int_0^t h(t') dt'\right) = \exp(-H(t)) ,
    \label{eqn:hazard}
\end{equation}
with $H(t)$ being a dimensionless quantity referred to as the \textit{cumulative hazard function}.
The cumulative hazard function can be interpreted as the expected number of failures up to a point if trials could be restarted at their point of failure.
The hazard function fundamentally encodes how failures occur within the system, whereas the survival function incorporates the statistical effects of how those failures are tested.

We follow the Weibull weakest-link model to describe surface-initiated breakdown. According to this model, the failure of a system comprised of many, independent elements is determined by the failure of its weakest element~\cite{weibull_statistical_1951}. In such case, the failure probability density function (PDF) as a function of electric field ($E$) can be expressed as:
\begin{equation}
   f_W(E;k,E_0,E_1) = \frac{k}{E_0} \left( \frac{E-E_1}{E_0} \right)^{k-1}\exp\left(-\left(\frac{E-E_1}{E_0}\right)^k\right) , 
    \label{eqn:weibull_pdf}
\end{equation}
where $k$, $E_0$, $E_1$ are the \textit{shape}, \textit{scale}, and \textit{location} parameters, respectively.
This is known as the three-parameter version of the \emph{Weibull PDF} and the corresponding \emph{Weibull cumulative hazard function} follows the power law 
\begin{equation}
    H(E) = \left(\frac{E - E_1}{E_0}\right)^k .
    \label{eqn:weibull_hazard}
\end{equation}
Taking $E_1$=0, we obtain the two-parameter version of the Weibull PDF.

%
%

%

As pointed out in Ref.~\cite{Hill_breakdown}, it is possible for the cumulative hazard function to be a function of both time and electric field, 
\begin{equation}
    H(t, E) = \left(\frac{t}{t_0}\right)^m\left(\frac{E}{E_0}\right)^n = \frac{E^{m+n}}{t_0^m\dot{E}^m  E_0^n},
    \label{eqn:hazard_time}
\end{equation}
where $m$ and $n$ encode the growth rate of hazards with time and field, respectively, and $\dot{E}$ is the ramp rate.
This shows that a scan over ramp speed is needed to determine the effect of previous breakdowns.
If $m=1$, the breakdown only depends on the field at the time of breakdown, while $m>1$ or $m<1$ demonstrate a progressive weakening or strengthening of the dielectric medium, respectively.


Lastly, it was mentioned in the previous section that electrode surface area is directly proportional to the number of asperities that could initiate breakdown.
A general property of weakest link models is that the effective hazard function of a system comprised of several elements with independent hazard functions will be the linear combination of those hazards. Hence, when the same electrodes are tested at different areas under the same conditions, one arrives at the \emph{area scaling formula},
\begin{equation}
    E_0 = C \cdot A^{-b},
    \label{eq:weibull_area_dependence}
\end{equation}
where $A$ is the SEA, $b=1/k$ is the area scaling exponent, and C is a constant factor. This formula predicts an inverse relationship between breakdown field and stressed area, which will be tested in this work.
The derivation of this formula can be found elsewhere~\cite{Tvrznikova:2018pcz,tvrznikova_direct_2019}.

\subsection{Field emission}
\label{sec:field_emission}

It has been suggested that breakdown does not occur in the liquid phase, but rather that streamers in the form of small, partially ionized gas bubbles form and then elongate along the field direction \cite{streamer, Phan:2020nhz}.
These streamers disrupt the electric field surrounding them, further enhancing the breakdown until a large, sustained current flows between the electrodes.
Field emission from the cathode provides both the initial injection of charge as well as the heat which grows the bubbles to the required length scale.

If this type of field emission is the primary driver of breakdowns, one can model the hazard as being proportional to the Fowler-Nordheim (FN) field emission current.
Applying this model to stainless steel electrodes, and following the convention from Ref.~\cite{wang:1997}, the current density ($J$) is correlated to the electric field in the bulk ($E$) as
\begin{equation}
\begin{aligned}
    J [\text{A}/\text{m}^2] = &\frac{1.54 \times 10^{-6}}{\phi} 10^{4.52 \phi ^{-1/2}} (\beta E)^2 \times \\ &\exp \left(\frac{-6.53 \times 10^9 \phi ^{3/2}}{\beta E}\right) ,
\end{aligned}
\end{equation}
where $\phi$ is the LXe stainless steel work function in eV, $E$ is given in \SI{}{V/m} and $\beta$ is the local field enhancement factor, such that $\beta E$ is the field at the tip of a field-enhancer (e.g.~an asperity).
Consequently, $\log(J/E^2)$ versus $1/E$, known as the FN plot, is expected to show a linear relationship with a negative slope. 


\subsection{The XeBrA experiment}
\label{sec:xebra}

\begin{figure}[htbp]
    \centering
    \subfloat{
        \includegraphics[width=0.45\columnwidth]{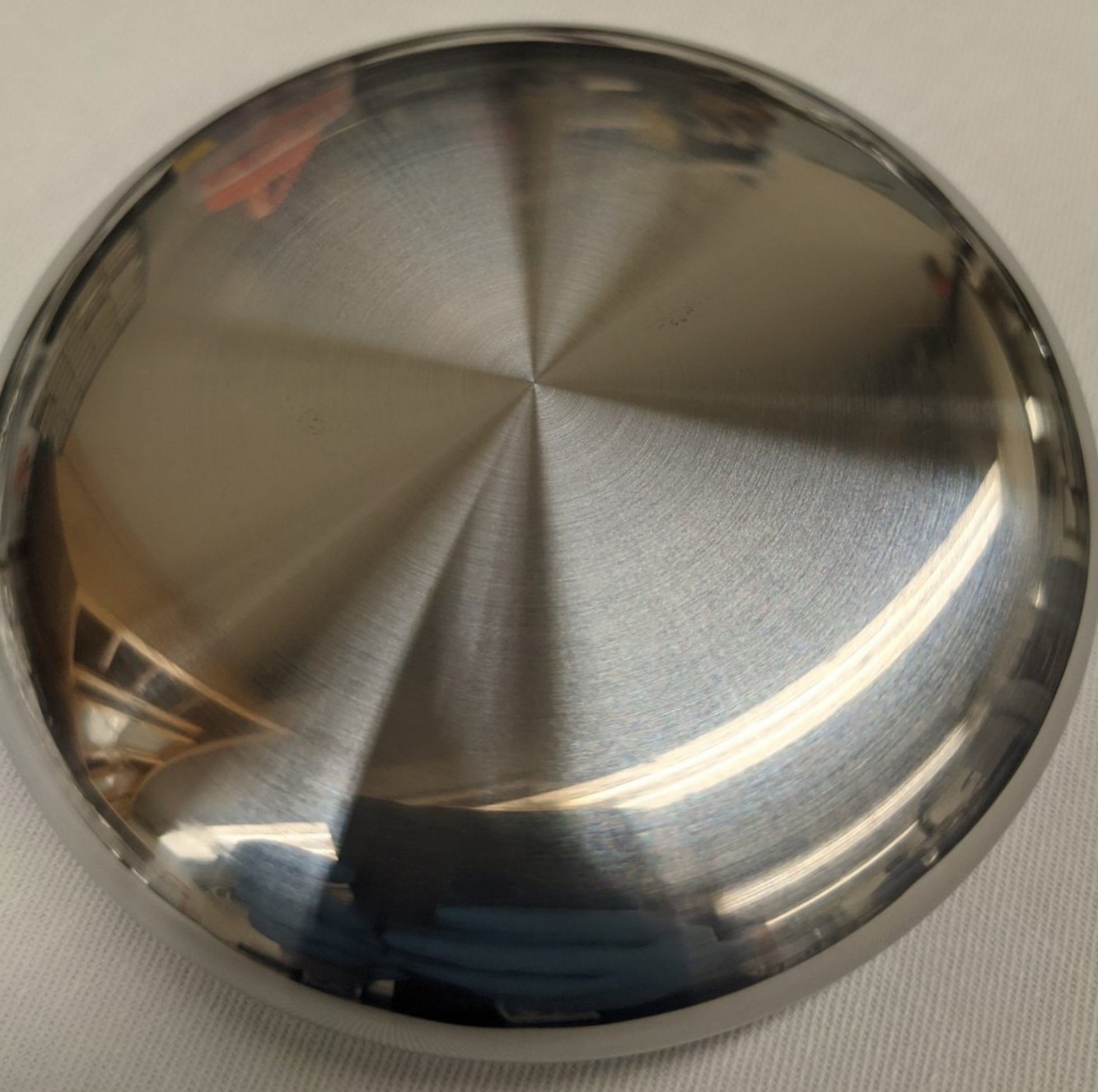}
    }
    \hfill
    \subfloat{
        \includegraphics[width=0.45\columnwidth]{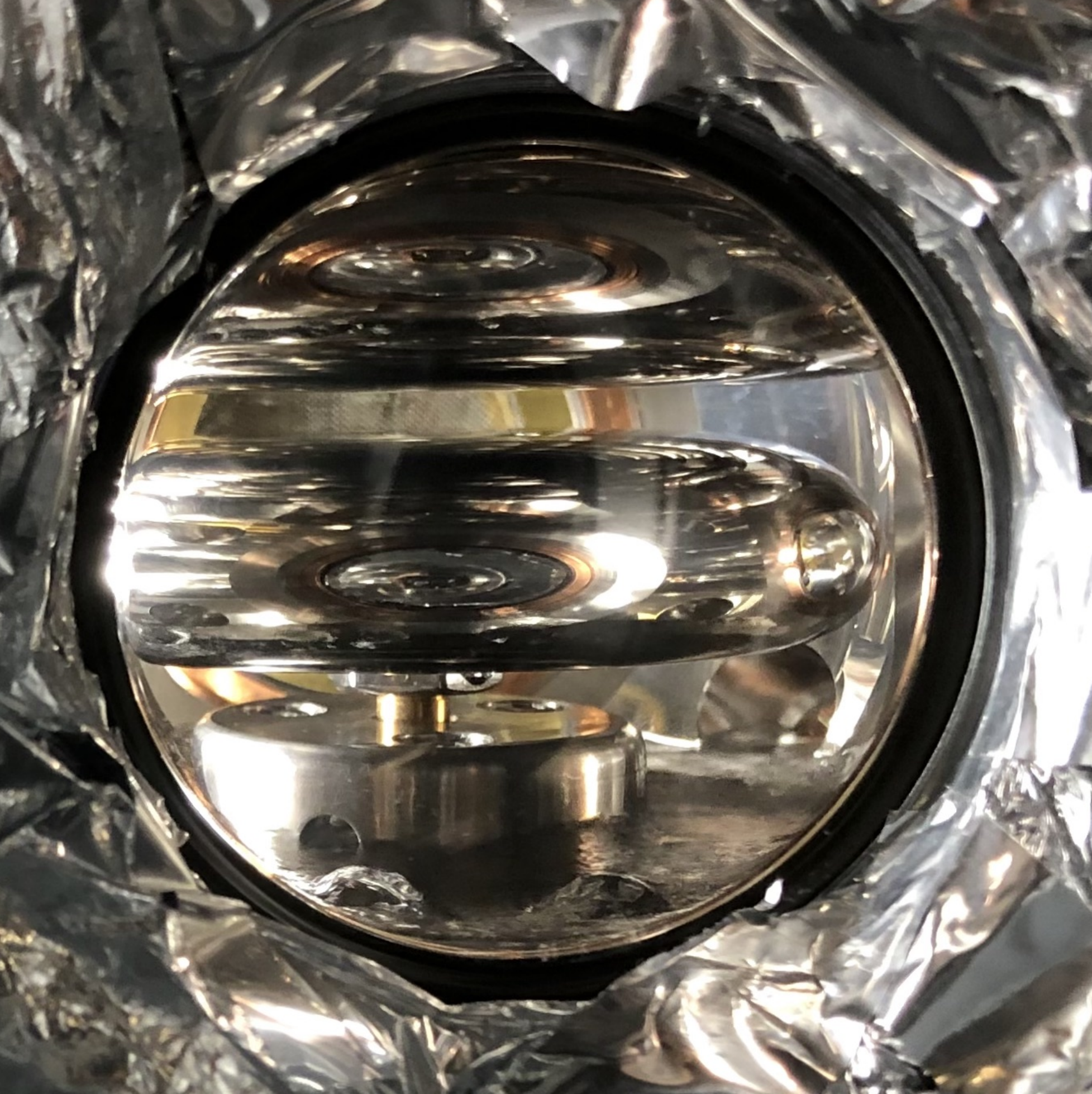}
    }
    \caption{\emph{Left:} A frontal view of one of the Rogowski electrodes.
    The surface was mechanically polished and cleaned before introducing it into the detector.
    \emph{Right:} A photograph of XeBrA's electrodes as seen through the front-side viewport during the filling of the chamber.
    Liquid xenon is visible below the cathode.}
    \label{fig:xebra}
\end{figure}

XeBrA is a five-liter stainless steel chamber with adjustable, large area electrodes that was built to investigate the factors affecting dielectric breakdown in noble liquids.
The electrodes, shown in \cref{fig:xebra}, have an axially symmetric Rogowski profile that generates highly uniform fields over a large area, resulting in a stressed electrode area that grows with gap distance~\cite{rogowski,Rogowski1923}. The anode is slightly larger ($r=56.6$ mm) than the cathode ($r=48.4$ mm) so that a uniform electric field between the electrons can be more easily achieved even when the two electrodes are not perfectly aligned.Three sets of electrodes were utilized in XeBrA, two of which were mechanically polished to four microinches, and the third underwent citric-acid passivation.

\begin{figure}
    \centering
    \includegraphics[width=\columnwidth]{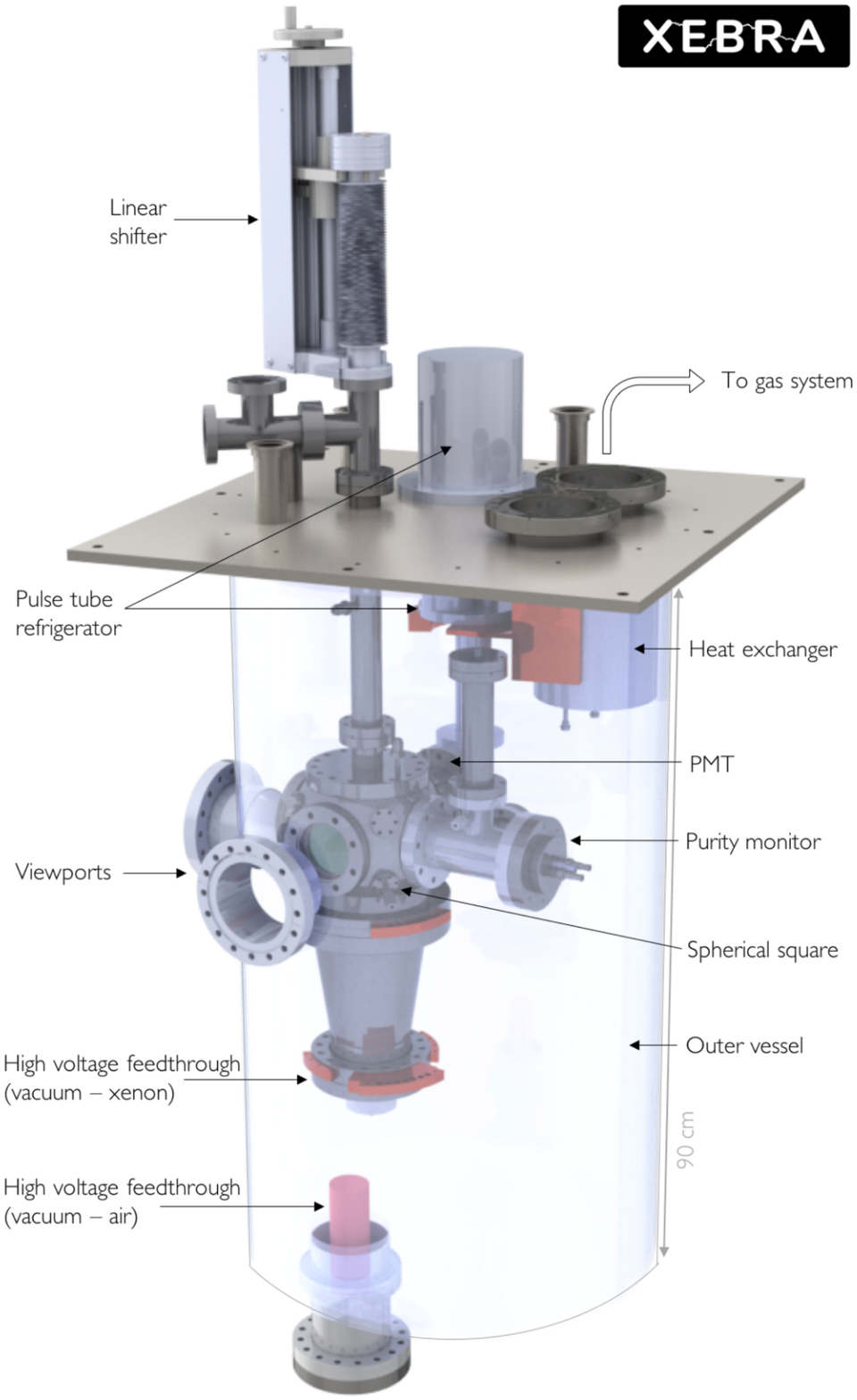}
    \caption{Diagram of the XeBrA system with its main components annotated (see text for details). Reproduced with permission from L.~Tvrznikova, \dquotes{Sub-GeV Dark Matter Searches and Electric Field Studies for the LUX and LZ Experiments,} Ph.D.~dissertation (Yale University, 2019)~\cite{Tvrznikova:2018pcz}.}
    \label{fig:xebra_design}
\end{figure}

A diagram of the XeBrA design is shown in~\cref{fig:xebra_design}.
The separation between electordes is controlled via a linear shift mechanism (LSM) with an affixed caliper for precise control.
Thus, a range of surface areas can be scanned by simply adjusting the electrode separation.
The cathode's orientation relative to the anode is adjusted by two set screws that are accessible through the viewports. 

A purity monitor (PM) measures the electron lifetime in-situ.
Low impurity levels are desirable in XeBrA, since the experiments XeBrA intends to inform typically operate with very long electron lifetimes, of the order of \SI{1}{ms}~\cite{LZ:2022ufs,XENONCollaboration:2022kmb,PandaX-4T:2021bab}. 
During a normal XeBrA run, gas circulates through a heated zirconium getter, which removes non-noble impurities to the part-per-billion (ppb) level. Furthermore, water and oxygen are removed from the xenon gas with an in-line purifier of 1 part-per-million outlet purity \cite{purifier_specs}.

The system is cooled through a heat exchanger (HX) coupled to a pulse tube refrigerator. 
The gas inlet pressure was controlled via a proportional-integral-differential (PID) loop operated by a programmable logic controller (PLC) that regulates the resistive heaters affixed to the heat exchanger. Both the high voltage ramping protocol and the resistive heaters were controlled via a LabVIEW program~\cite{bitter2006labview}. 
It is worth emphasizing that the temperature is not controlled independently of the pressure.
Instead, the stainless steel in XeBrA is typically cooled below the boiling curve in order to suppress bubbles.
The circulation speed was adjusted to tune the HX efficiency as necessary. A diagram of the piping and instrumentation is shown in \cref{appd:xebra_pid} for extra reference. 

A unique aspect of XeBrA is the presence of two viewports that frame the electrodes,
allowing visual confirmation that the liquid is in a bubble-free state.
Moreover, a pair of high-speed cameras\footnote{AOS Technologies PROMON u750 cameras} are installed in each of the viewports.
The cameras are held in place with a 3D-printed mount and housing and they record at up to 2,146 frames per second. 
This framerate allows each camera to capture at least one frame of the breakdown per typical breakdown event.
The two-camera system serves two main purposes: first, to monitor the presence of bubbles in the liquid, and second, to reconstruct the position of the breakdown in between the electrodes (see~\cref{sec:pos_recons}). More information about the XeBrA experiment can be found in Refs.~\cite{tvrznikova_direct_2019,Tvrznikova:2018pcz}.

\section{Methods}
\label{sec:methods}
\subsection{Experimental procedure}
\label{procedures}

The operating loop of XeBrA consists of the cathode HV being gradually increased until the current through the High Voltage Power Supply (HVPS) exceeds a set threshold (typically between 1 and \SI{10}{\uA}).
This causes a fault signal to be sent by the power supply and received by the control software, which then automatically ramps the voltage down to zero. 
The control software then waits a prescribed amount of time before initiating the next ramp.
During this process, the HVPS voltage, current, system pressure, temperature, and circulation speed are logged at a frequency of \SI{10}{Hz}. The current into the anode is read out via a charge-sensitive preamplifier and Gaussian shaping amplifier combination.

The data-taking procedure for XeBrA consists of an eight-day cycle during which xenon is condensed, circulated, and recovered back into a high-pressure aluminum bottle. 
Each cycle is labeled with a Run number and generally contains three complete workdays of breakdown data. Two bar of gaseous xenon are initially introduced into the chamber, while the stainless steel cools down to xenon's boiling point.
The condensation procedure commences following two days of circulation at 1.2--1.4 bar absolute.
Approximately 12 kg of xenon are condensed on the first day, with the fill path passing through an in-line chemical purifier.
The xenon is then circulated through a getter\footnote{PF4-C15-R-1, manufactured by SAES Pure Gas} overnight for purification at 1.4 bar absolute.
Immediately prior to the first day of breakdown data-taking, an additional 4 kg of xenon are condensed via the circulation path, raising the pressure to 2 bar absolute.
This additional condensation rapidly raises the condensation temperature, suppressing bubble formation near the cathode high voltage (CHV) feedthrough.

In order to further reduce the risk of bubble formation, a \dquotes{subcooling procedure} was followed during the nights between data-taking.
This procedure consisted of lowering the set point pressure from 2.0 to 1.3--1.4 bar absolute, while lowering the temperature on the heaters located at the gas inlet ports.
This cools the stainless steel along the boiling curve of xenon, and thermal equilibrium is reached after several hours.
The pressure is rapidly raised the following day by reactivating the inlet heaters and increasing the speed of the circulation pump.
After the gas pressure increases to the desired set point, a temperature inversion is established, with the bottom of the chamber being colder than the top.
The bulk of liquid xenon is then sufficiently far away from the boiling curve to suppress bubble nucleation on the CHV feedthrough.
It was observed that this temperature inversion induced a placid state that could be maintained for up to 8 hours, as verified by visual inspection via the viewports.

The light produced during electrostatic discharge (ESD) is detected in several ways.
During all the runs except the first one (i.e.~Runs 5--10), breakdowns were observed with a pair of high--speed cameras, oriented approximately perpendicular to each other, enabling position reconstruction of the breakdown location. Cameras were not installed for Runs 1--3, corresponding to the measurements that were presented in Ref.~\cite{tvrznikova_direct_2019}, and they were not fully operative for Run 4.
During Runs 5--8, the cameras were occasionally replaced with a ONSemi 60035C-Series 6x6 mm Silicon Photomultiplier (SiPM)~\cite{sipm} in order to record the onset of electroluminescence.
For Runs 9--10, an additional feedthrough was installed which enabled simultaneous camera and SiPM operation. 

Data were also taken with fixed voltage in order to investigate the time evolution of the single-photon rate.
Rather than ramping until breakdown, the cathode HV is increased to \SI{30}{kV} and maintained there for 10 minutes.
At a separation of \SI{5}{mm}, it was determined that the risk of breakdown was negligible at such voltage, and no discharges were observed in this data-taking mode.
Photons were detected with a Photomultiplier Tube (PMT) model R9228 from Hamamatsu.

The XeBrA electrodes are negatively polarized, with the cathode receiving negative high voltage and the anode connected to ground indirectly through a charge-sensing electronics chain.
Two surge protection circuits were placed between the  anode connection itself and a charge sensitive preamplifier.
The first was a spark gap for large surges and was placed near the feedthrough into the inner chamber, and the second was a resistor-TVS diode circuit for smaller surges and was placed before the amplifier.
A 50 ohm coaxial cable connected the circuits.
The preamplifier was AC coupled to the grounding circuit, allowing it to sense the fast signal from breakdowns and other phenomena.

The amplification board consisted of a Cremat CR-150 charge sensitive amplification board and CR-210 shaping amplifier board, with CR-111 and CR-200-\SI{1}{\us} amplification chips, providing Gaussian pulse response with a full-width half maximum (FWHM) of \SI{2.4}{\us}.
The gain of the shaping amplifier could be continuously adjusted using a potentiometer, and as such a square wave from a function generator was used to calibrate the overall gain of the complete electronics chain.
This signal was digitized with a data acquisition system (DAQ) from National Instruments with a sample rate of 3.5 MHz, allowing for zero dead time acquisitions. The Polaris software~\cite{Suerfu_2018} was employed to facilitate writing DAQ data to disk.

Finally, it is worth highlighting the data structure for the analysis that is presented. In order of least to most granular, the data collected in XeBrA can be organized as follows: run, dataset, breakdown, and event. Runs indicate a cycle of LXe filling and recovery, datasets indicate a particular set of data-taking conditions (e.g.~electrode gap separation and pressure), breakdowns (or alternatively, ramps) indicate a singular high voltage ramp from \SI{0}{V} until discharge occurs, and events denote a division of the digitized data, typically \SI{280}{ms} long.



\subsection{Cathode tilt angle}
\label{sec:tilt_estimate}


Whenever the electrodes were manipulated in some way (e.g.~cleaning or exchanging them), the tilt angle and LSM caliper tare were set. 
The separation was calibrated by placing a shed-free wipe between the electrodes and squeezing them together.
The wipe's thickness was subtracted from the caliper reading to obtain the reading that would result in a crash. 
A photogrammetric approach was utilized to estimate the angle of the cathode plane relative to the anode plane.
This tilt angle has important consequences for the data analysis described in the next sections, so a precise estimation was critical.
Photographs of the electrodes were taken from each viewport with a high-resolution camera at known electrode separation gaps.
In addition, a couple of photographs were taken at separation distances unknown to the analyzer, which later would be used to validate the accuracy of the method.

The tilt estimation procedure considers eight locations across the electrodes for a given image.
The LSM caliper readings which would result in the electrodes being in contact with one another at those locations were determined.
These values were found by recording the separation distance in pixels between the electrodes at each of the eight horizontal positions across different photographs at various separation gaps. Then, for a given location an orthogonal distance regression~\cite{Boggs1987ODR} (ODR) was applied to the separation gap distance as a function of the pixel separation and the y-intercept from the fit was interpreted as the caliper reading for the electrodes being in contact at that location.



After this process was repeated across the chosen locations, a linear regression of the y-intercepts as a function of real-space distances was used to find the tilt.
The dimensions of a reference object in the frame were used to scale the number of pixels into a physical distance.
The radius of the cathode is known to be 48.4 $\pm$ 0.1 mm and the number of pixels that make up the length of the cathode can be estimated.
The precision of this conversion factor was found by applying it to the images taken at blinded separations.
The average difference between the determined separation value found using this conversion and the known value was 0.5 mm.
An additional ODR fit was then performed to the y-intercepts as a function of real-space distances where two sources of systematic error were considered: uncertainty in the measurement of the cathode (0.1 mm) and uncertainty in counting pixels (0.5 mm).
The slope from the fit, along with the uncertainty on that slope, gives a tilt and its uncertainty.
These angles are shown in \cref{tab:tilts} for all the Runs that were conducted in this work.


\begin{table}[htbp]
    \centering
    \begin{tabular}{|c|c|c|c|}
    \hline
    Runs & Front  [$^\circ$] & Side  [$^\circ$] & Total   [$^\circ$] \\
    \hline
         4-5 & \: 0.1 $\pm$ 0.1  & -0.6 $\pm$ 0.1 & 0.6 $\pm$ 0.1\\
         6-7 & -0.2 $\pm$ 0.1 & -0.2 $\pm$ 0.1& 0.3 $\pm$ 0.1\\
         8 & \: 0.3 $\pm$ 0.1 & -1.0 $\pm$ 0.1 & 1.0 $\pm$ 0.1\\
         9 & \: 1.2 $\pm$ 0.2 & -0.4 $\pm$ 0.1 & 1.3 $\pm$ 0.1\\
         10 & -0.1 $\pm$ 0.1 & \: 0.5 $\pm$ 0.1& 0.5 $\pm$ 0.1\\
         \hline
    \end{tabular}
    \caption{Estimated cathode tilt angles for different Runs.
    The front and side tilt labels refer to the tilts of the electrodes in the image plane for images taken from the side and front viewports, respectively, and the total value is obtained from adding those angles in quadrature.}
    \label{tab:tilts}
\end{table}

\subsection{Breakdown position reconstruction}
\label{position_reconstruction}

In a typical data-taking run, hundreds of videos containing breakdowns were recorded.
The still frames that contain the bright flash from the streamer are extracted for further analysis by applying an overall brightness filter. 
Each of the video frames passing the filter was reviewed to ensure that the discharge is in the frame. The frames that contain a discharge are undistorted using OpenCV-python~\cite{opencv_library}.

The breakdown-containing frames from the individual cameras are paired using the image timestamps. 
The location of the breakdown in pixel space was recorded for each frame, corresponding to the brightest vertical strip of the frame.
The pixel coordinates of the breakdown are then transformed into the real-space location relative to the cathode surface following a two-step method.
First, a projective transform was applied. The parameters for this transform were found by using four fixed known points in pixel space and their corresponding exact real space locations at the edges of the bottom electrode. The result of this transform is the projection of two-pixel space coordinates, coming from nearly orthogonal axes, onto the surface of the electrode.
Second, a 2nd-order polynomial transformation was applied to accurately determine the location of the breakdown in XY space.
The parameters that are used in this second transformation were found by minimizing the residuals between positions found after the projective transform and all corresponding known positions in a calibration dataset. For this calibration dataset, 26 images of a target with known geometry were taken in a test setup. In the setup, the external can of XeBrA was removed and the target was placed at an equivalent location as the cathode inside the can.



\begin{figure}[htbp]
    \centering
    \includegraphics[width=0.9\columnwidth]{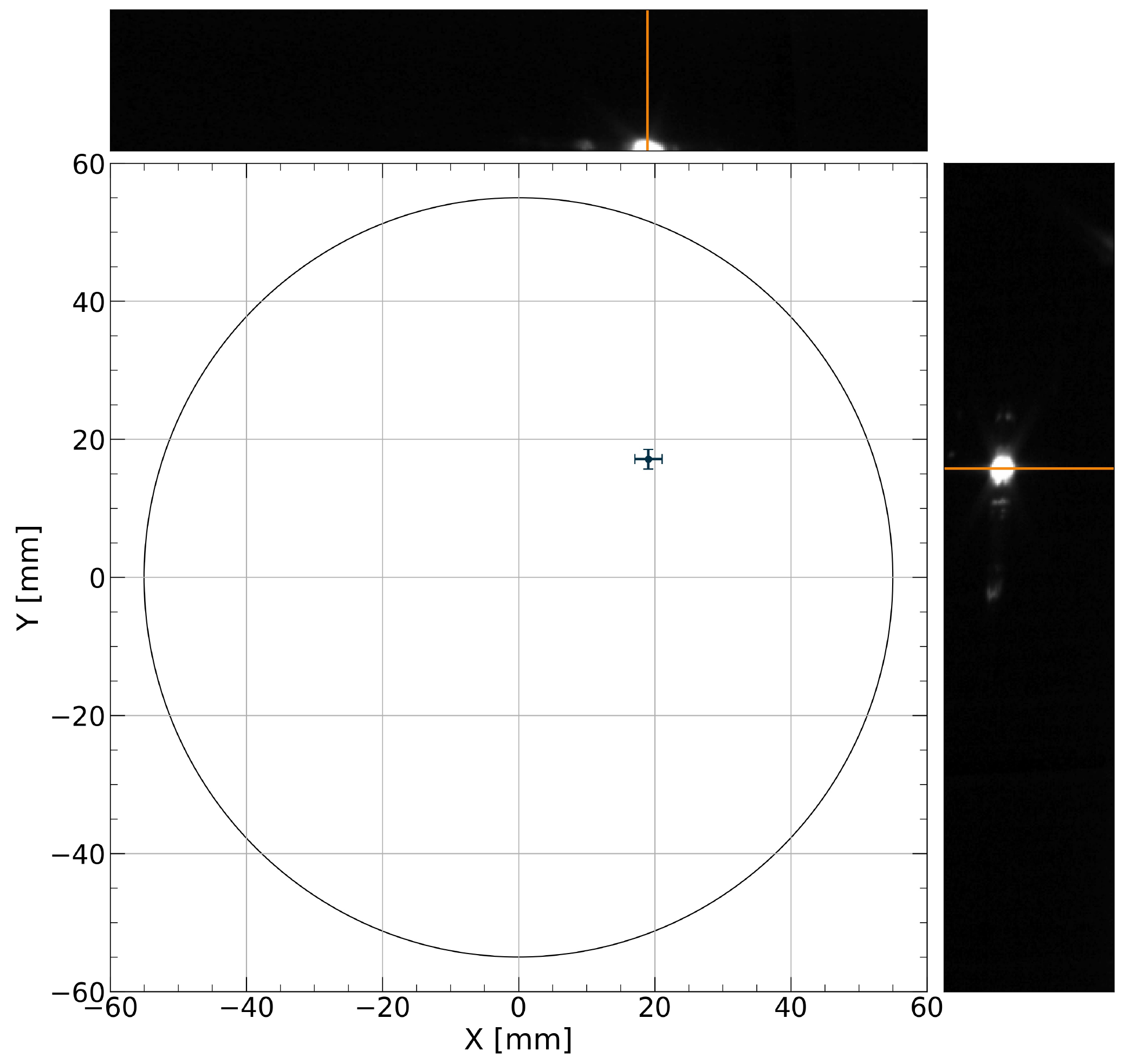}
    \caption{Illustration of the position reconstruction analysis. The top frame corresponds to the frontal camera and the right frame to the side camera. The brightest regions in each frame (marked with a solid orange line) are used as input in the transformation from pixel space to real space, resulting in the point shown in the XY plane. The error bars reflect the systematic uncertainty on the location of the breakdown in pixel space.}
    \label{fig:posreco_ex}
\end{figure}

The result of this procedure is the real-space locations of the breakdowns on the electrode's surface. As an example, \cref{fig:posreco_ex} shows an illustration of a reconstructed breakdown in XY space along with the camera frames that were used in the reconstruction. A solid orange line marks the location of the brightest region in each frame, which is then used as input for the two-step transformation. The error bars shown in that figure are calculated by propagating the error on the location of the breakdown in the frames through the two steps in the position reconstruction procedure.

\subsection{Electric field simulation}
\label{sec:E_sims}

\begin{figure}[htbp]
    \centering
    \includegraphics[width=1\columnwidth]{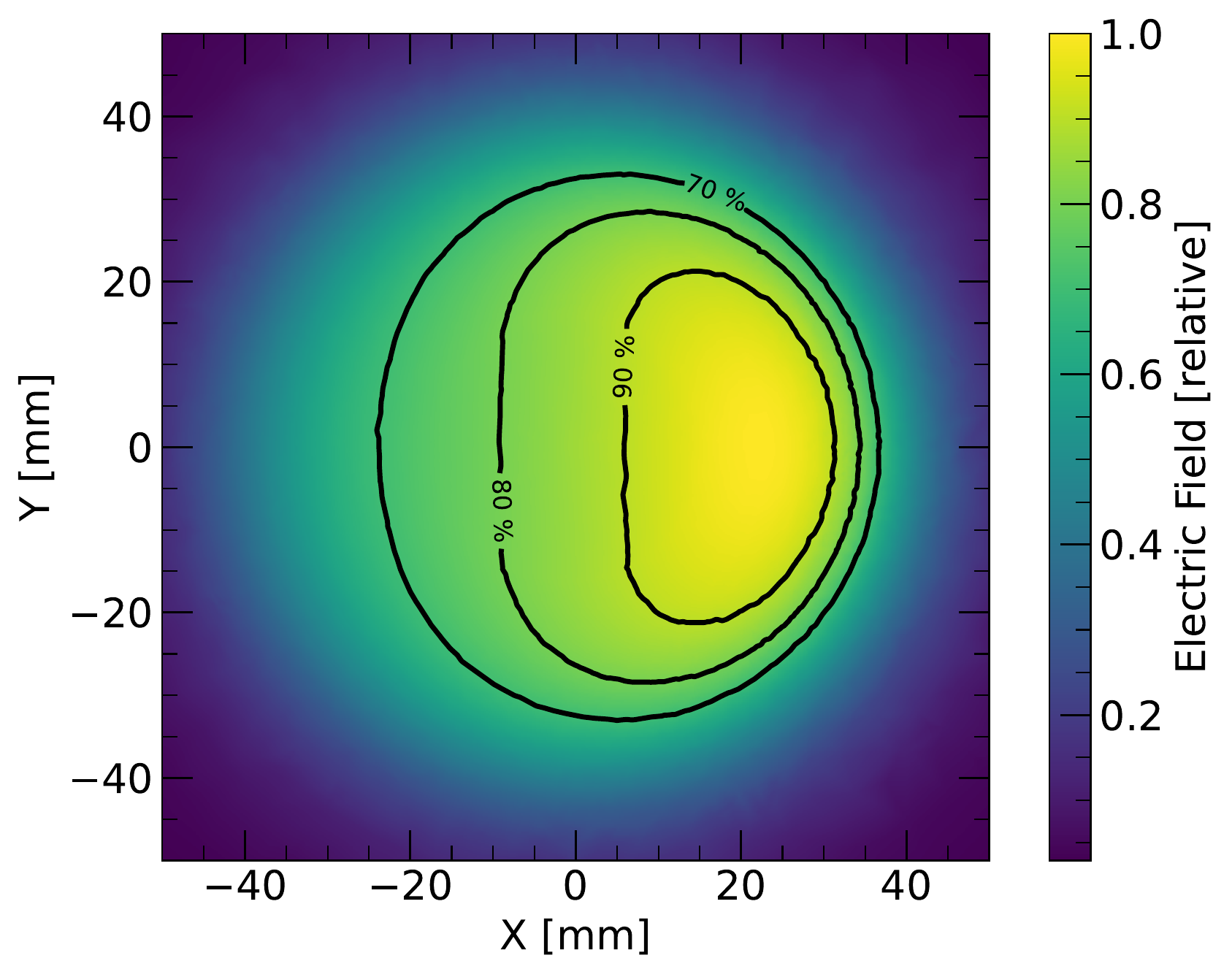}
    \caption{Electric field magnitude above the cathode for a \SI{3}{mm} electrode separation and 1.3$^\circ$ cathode-anode tilt.
    Isocontours are indicated in black and the color map shows the electric field distribution normalized by its peak value.
    The SEA is defined as the region with an electric field magnitude equal or greater than 90\% of the peak field.
    }
    \label{fig:field_cathode}
\end{figure}

The finite element simulation software FEniCS~\cite{fenics} was used to estimate the stressed area at different electrode separations and cathode tilt angles.
In the following, we adopt the subsequent definition of the SEA: the area of the electrode surface with an electric field magnitude equal or greater than 90\% of the maximum electric field.
The field is simulated on a grid of separations and cathode tilt angles (see an example in~\cref{fig:field_cathode}).
Note that the SEA is a monotonically increasing function of separation distance, growing from \SI{11}{cm \textsuperscript 2} at \SI{1}{mm} to \SI{30}{cm \textsuperscript 2} at \SI{5}{mm} (at zero tilt). 

After calculating the SEA and maximum field magnitudes, the stressed area is interpolated for the angles inferred by the photogrammetric approach described in \cref{sec:tilt_estimate}.
The tilt introduces a correlated uncertainty on the calculation of the SEA at all separations.
To find the uncertainty on the SEA, a toy Monte-Carlo (MC) simulation is run, whereby several thousand cathode tilt angles are sampled from a Gaussian distribution with mean and standard deviations taken from~\cref{tab:tilts}.
At each MC point, the SEA is interpolated from the finite element solutions.
The uncertainty on SEA is then taken as the standard deviation of the sampled SEA values for a given configuration.
The resulting uncertainties are tabulated in~\cref{tab:sea_uncertainy}.

\begin{table}[htbp]
    \centering
    \begin{tabular}{|c|c|cc|}
    \hline
    Runs & Tilt [$^\circ$] &   \multicolumn{2}{c|}{SEA [cm$^2$]}  \\
    \hline
    {} & {} & \multicolumn{2}{c|}{Electrode separation} \\
    
    {} & {} & \SI{1}{mm} & \SI{3}{mm} \\
    \hline
         4--5 & 0.6 $\pm$ 0.1  & 6.0 $\pm$ 0.7 & 16.8 $\pm$ 1.1\\
         6--7 & 0.3 $\pm$ 0.1 & 9.6 $\pm$ 1.5 & 19.7 $\pm$ 0.8\\
         8 & 1.0 $\pm$ 0.1 & 4.0 $\pm$ 0.3 & 12.3 $\pm$ 0.9\\
         9 & 1.3 $\pm$ 0.1 & 3.5 $\pm$ 0.4 & 10.7 $\pm$ 1.3\\
         10 & 0.5 $\pm$ 0.1 & 6.8 $\pm$ 1.0 & 17.8 $\pm$ 1.0 \\
         \hline
    \end{tabular}
    \caption{Estimated SEA and uncertainty for each Run at two representative separations. The effect of the tilt uncertainty is more notable in the \SI{1}{mm} electrode separation.}
    \label{tab:sea_uncertainy}
\end{table}

%


\subsection{Purity monitor}
\label{sec:Purity}
The concentration of oxygen-equivalent impurities contained in the LXe were measured by an in-situ purity monitor. The monitor measures the electron lifetime as they drift between a gold-plated cathode and a brass anode. The electrons are generated via the photoelectric effect by a xenon flash lamp, with UV photons guided to the cathode by a fiber optic cable. Frisch grids allow for the cathode signal to be distinguished from the anode signal. The pulses are read out by a Cremat CR-111 charge amplifier and CR-200-\SI{1}{\us} shaping amplifier, generating easily distinguishable Gaussian pulses. Data are digitized using a oscillosope and analyzed offline.
Then, the electron lifetime is found by filtering the voltage signal with a boxcar filter and taking the ratio of the corresponding pulse heights.
This method also yields highly accurate timing information for the pulses themselves.
The ratio is converted into lifetime as
\begin{equation}
    \tau = \frac{t}{\log(A_C/A_A)} ,
    \label{eqn:lifetime}
\end{equation}
where $t$ is drift time, $A_C$ is the cathode pulse area and $A_A$ is the anode pulse area. The oxygen-equivalent impurity is calculated utilizing the attachment coefficients found in Ref.~\cite{bakale1976impurity_attachment}, which results in the conversion formula
\begin{equation}
{O_2}  = \frac{455 \text{ ppb} \cdot \SI{}{\us}}{\tau}.
\label{eqn:purity}
\end{equation}

The oxygen-equivalent impurity measurements for the 2020--21 data-taking runs are shown in \cref{tab:purity}. 
During Run 5, an overpressure incident led to xenon mixing with residual gas in one of the recovery vessels.
The purity of the xenon after this incident was estimated using a sampling system and it was found to be less than 100 ppb. To estimate the purity, we followed a cold trap method similar to the one described in Ref.~\cite{Leonard_2010}.
Run 6 was aborted before a purity measurement could be taken and Run 7 suffered a data corruption issue.

The uncertainties on the purity monitor pulse areas are found by examining the pre-cathode pulse signal RMS, while the uncertainty on the drift time is taken to be twice the sampling period of the oscilloscope.
The overall results are tabulated in \cref{tab:purity}.
The discrepancy in uncertainties is primarily due to inconsistencies in the number of pulses the oscilloscope averaged over.

\begin{table}[htbp]
    \centering
    \begin{tabular}{|c|l|l|}
    \hline
        Run &  Electron lifetime [\SI{}{\us}] &  O$_2$-equivalent impurity [ppb] \\
       \hline
        4 & 19.4480 $\pm$ 0.0007 & 23.3470 $\pm$ 0.0008 \\
        5-7 & - &  <100 \\
        8 & 34.1 $\pm$ 0.1 &13.34 $\pm$ 0.05 \\
        9 & 77 $\pm$ 2 & 5.9 $\pm$ 0.1 \\
        10 & 125 $\pm$ 5 & 3.6 $\pm$ 0.2 \\
        \hline
    \end{tabular}
    \caption{Purity measurements taken during the 2020--21 data-taking runs.
     A general trend of lower purities are observed between consecutive runs.}
    \label{tab:purity}
\end{table}

\section{\label{sec:Results} Results}

\subsection{Stressed electrode area scaling}
\label{sec:Stressed Area}

At each separation and run dataset, the Weibull parameters described in \cref{sec:hazard_functions} were extracted.
The breakdowns were initially filtered to remove confounding data.
The selection criteria were: 1) breakdown voltage must be larger than 4 kV due to occasional spurious trips on the ramp start; 2) pressure must be within \SI{0.05}{bar} of the set point; 3) the first ten breakdowns of any dataset are vetoed to mitigate possible conditioning effects; 4) the breakdown must occur at least \SI{0.1}{s} away from a step in voltage, which ensures that we only consider data during periods of quasi-static cathode voltage. 

\begin{figure}[htbp]
\centering
    \includegraphics[scale =0.6]{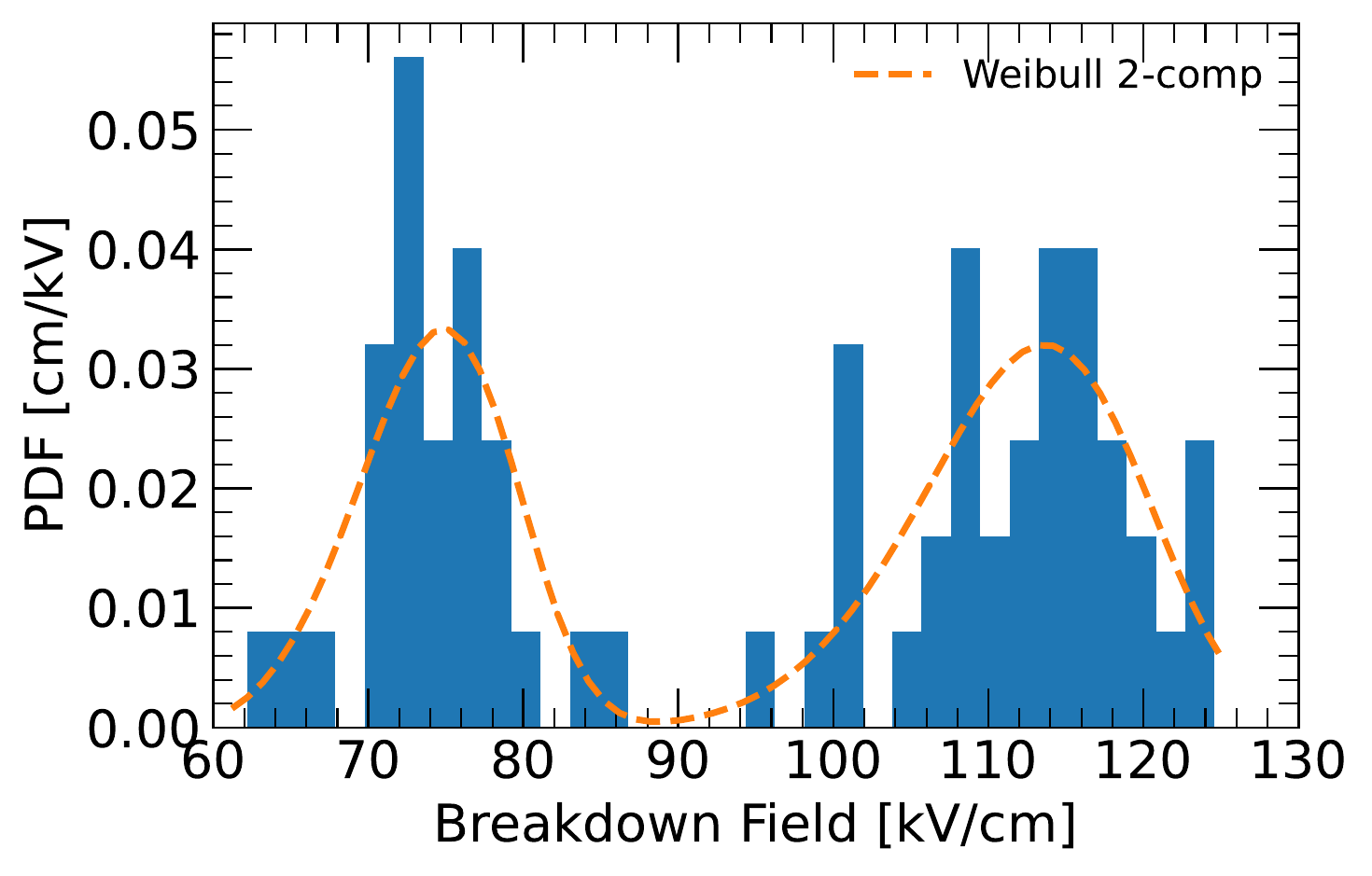}
    \qquad
    \centering
    \includegraphics[scale =0.6]{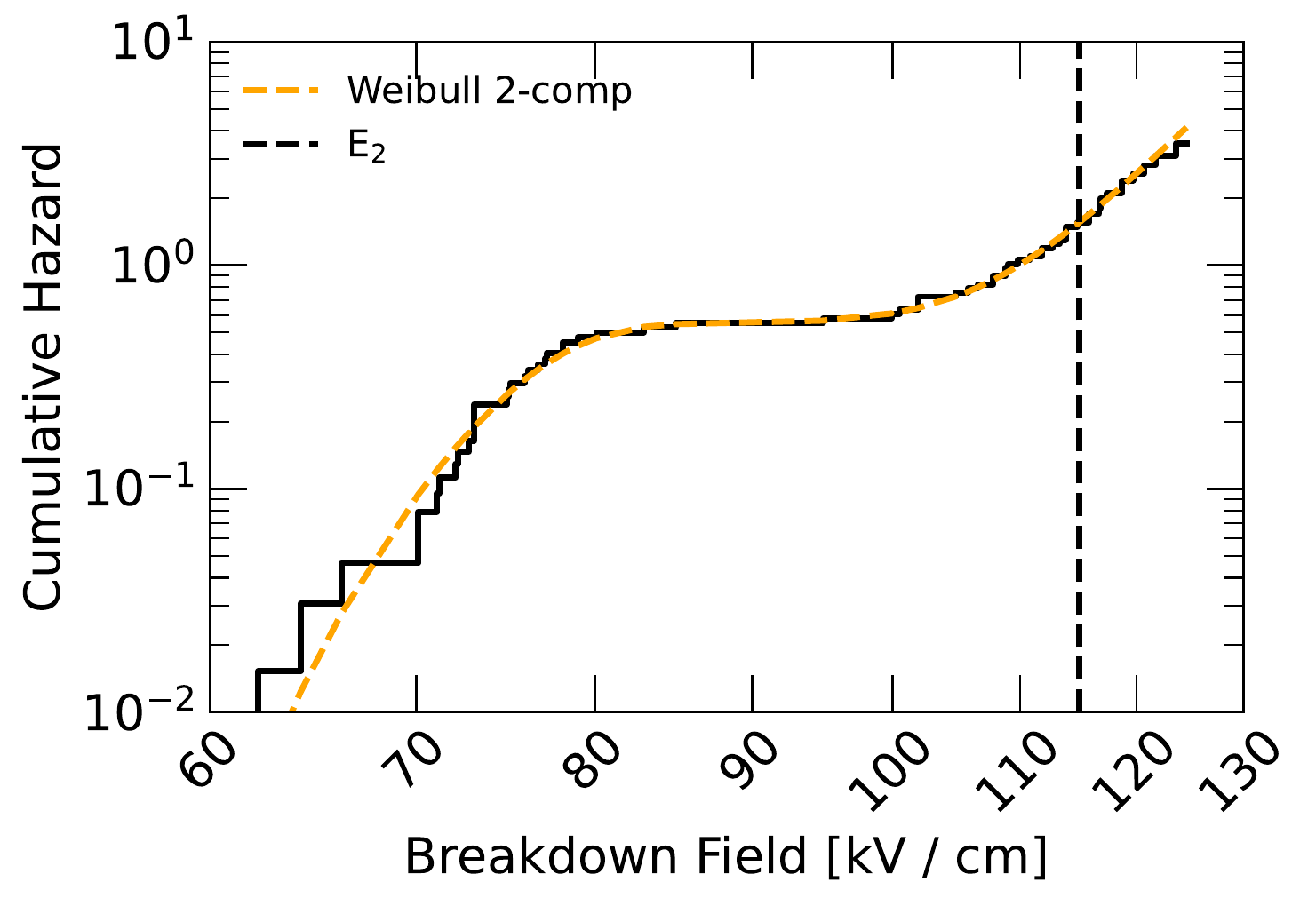}
 \caption{An example of a dataset that strongly favors a two component, three-parameter Weibull PDF (i.e.~Model 3, see text). This is for Run 4 and electrode separation 1 mm. \emph{Top:} Histogram of breakdown fields (blue) and two-parameter Weibull PDF overlaid on top. \emph{Bottom:} The cumulative hazard distribution (black) and the model that provided the best fit to data (orange dash line). $E_2$ is equal to the shifted scale parameter $E_0+E_1$, corresponding to the 63rd quantile of the distribution, and is plotted here for reference (black dash). The cumulative hazard can be interpreted as the expected number of breakdowns that need to occur before reaching a given breakdown field, if ramps were restarted at the point of failure.
 }
    \label{fig:weibull_2comp}
\end{figure}

The top panel of \cref{fig:weibull_2comp} shows a histogram of breakdown fields for an electrode separation of \SI{1}{mm} obtained in the first run of this data campaign (Run 4). The breakdown field was calculated as the breakdown voltage divided by the simulated electrode separation at the maximum electric field (see~\cref{sec:E_sims}). Two distributions are apparent, one at lower fields than the other, which is indicative of transient dips in electrode performance. This feature of the data is represented by marked corners in the cumulative hazard distribution, shown in the bottom panel and obtained using the Kaplan-Meier estimator~\cite{doi:10.1080/01621459.1958.10501452}. Other datasets showed a single distribution of fields, resulting in a more smooth cumulative hazard distribution. Given this variability in the data collected, we tested three different models for each dataset:
\begin{enumerate}
    \item One component, two-parameter Weibull PDF
    \item One component, three-parameter Weibull PDF
    \item Two components, three-parameter Weibull PDF
\end{enumerate}
Model 3 is constructed as a normalized linear combination of two Weibull PDFs.
Model 1 is treated as the null hypothesis initially, and against it the likelihood ratio of Models 2 is tested: 
\begin{equation}
\begin{aligned}
    &\ln\left(\frac{\mathcal{L}_2}{\mathcal{L}_1}\right) =\\
    &\sum_{i=1}^N \left(\ln f_W(E_i; k^{(2)}, E_0^{(2)}, E_1^{(2)}) 
    -\ln f_W(E_i; k^{(1)}, E_0^{(1)})\right).
\end{aligned}
\label{eq:likelihood_ratio}
\end{equation}
The subscript $i$ runs over data points while the superscript indicates the corresponding Weibull parameters of each model.
Model 1 is rejected if the $p$-value of this likelihood ratio scores below 5\%.
In that case, the process is repeated for Model 2 and Model 3, with Model 2 as the null hypothesis. With this method we test if progressively adding more parameters to the model improves the fit to data significantly.
If Model 3 is selected, the best-fit parameters from the larger component are chosen for further analysis.
The reason for this choice is that we believe the best-fit parameters from the higher component are more representative of the breakdown mechanism under study. The lower component in those fits are most likely due to temporary instabilities of the system caused by the presence of debris or bubbles, transient changes in the thermodynamics, or surface pitting. These low breakdown field events were clustered in specific periods of time, suggesting that they are not a symptom of some underlying condition.

Large individual uncertainties on the scale ($E_0$) and location ($E_1$) parameters are present when Model 2 or 3 were selected. 
Also, a high degree of anti-correlation between those two parameters was observed.
This is expected, since statistical quantities such as the mean and variance of the Weibull PDF depend on all of the model parameters. 
Combinations of those two parameters were used to provide more statistically robust and interpretable metrics.
In particular, the shifted scale parameter $E_2 \equiv E_0+E_1$ was used since it has approximately the same interpretation as the scale parameter of a two-parameter Weibull PDF (i.e.~the 63rd percentile of the distribution).
One consequence of the large anti-correlation observed between the $E_0$ and $E_1$ parameters is that the off-diagonal coefficients of the covariance matrix become sufficiently large to play a relevant role in the propagation of uncertainty, resulting in a lower uncertainty overall for the shifted scale parameter $E_2$. The best-fit parameters for all the datasets taken in this campaign are shown in the appendix (see \cref{tab:run_fits}).


An alternative approach to fit the data would be to use a linear combination of cumulative hazard functions, $H(E)$, but that was observed to return generally worse fits for the datasets where deviations from linearity were seen.
This can be interpreted as indicating that a transient hazard was observed, rather than simultaneous hazards (e.g.~the flat region shown in the bottom panel of \cref{fig:weibull_2comp} cannot be modelled with a linear combination of cumulative hazard functions.)

%
%

%

\begin{figure}[htbp]
\centering
   
    \includegraphics[width = .95\columnwidth]{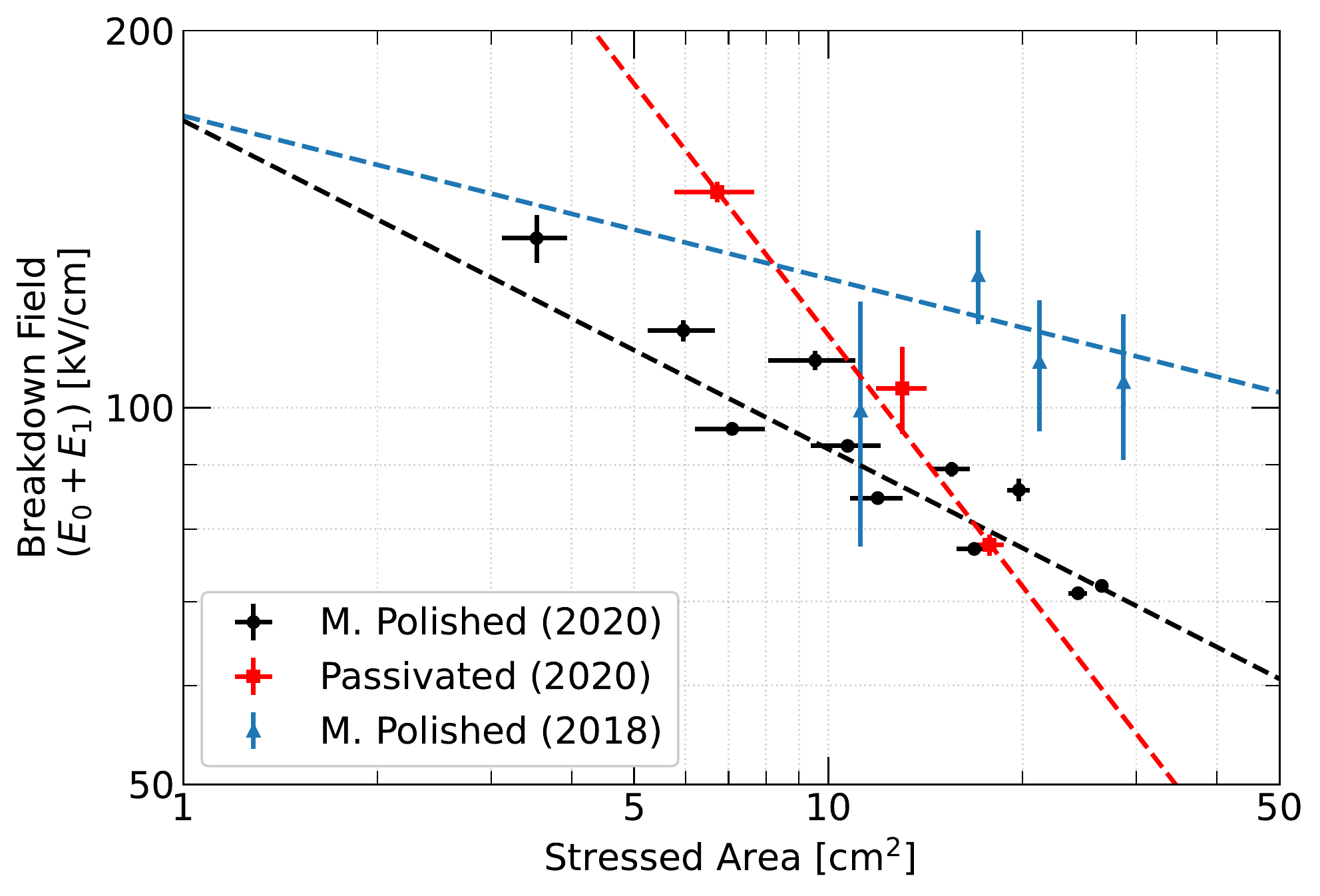} \qquad
     \includegraphics[width=0.95\columnwidth]{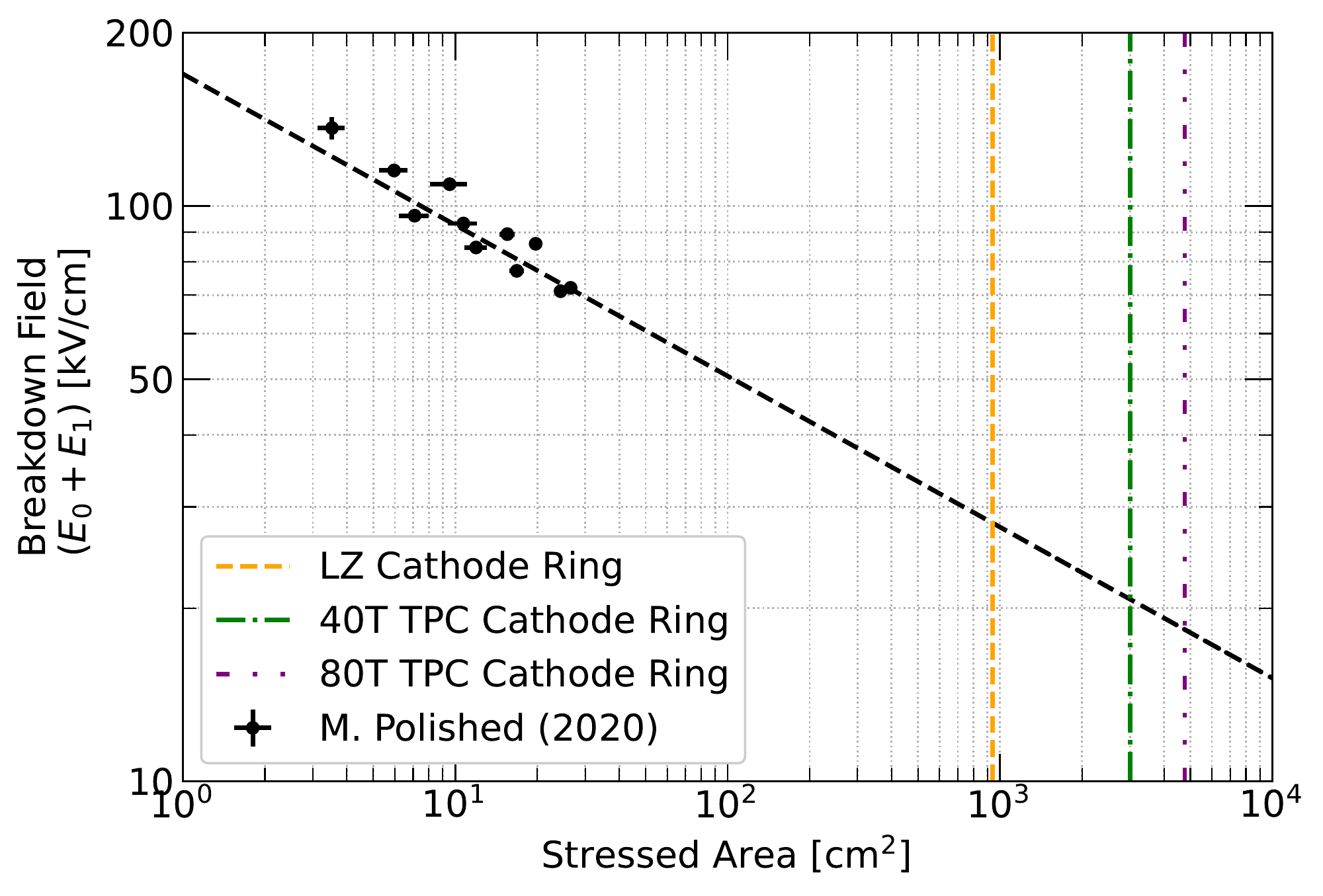} \qquad
    \caption{
    The dependence of breakdown field on stressed area. 
    The Weibull scale plus location parameters are plotted ($E_2  \equiv E_0 + E_1$).
    This statistic was chosen due to the high level of correlation between the scale ($E_0$) and location ($E_1$) parameters.
    \textit{Top}: The current stressed area scaling using mechanically polished (black) and passivated electrodes (red) are compared to data from the 2018 analysis (blue), in which mechanically polished electrodes were used.
    Although the data points are taken at identical separation distances, there are more stressed areas displayed in the 2020 measurements due to the different estimated cathode tilt angles between some runs. 
    The blue data points were reproduced with permission from JINST 14 P12018 (2019)~\cite{tvrznikova_direct_2019}. Copyright IOP Publishing, Ltd.
    \textit{Bottom}: The 2020 mechanically polished data and polynomial fit extrapolated to larger SEA values. In addition, the estimated SEAs for LZ (7 tonnes active LXe), and two possible next-generation TPCs (with 40 and 80 tonnes active LXe, respectively) are shown.
    }
    \label{fig:sea_scaling}
\end{figure}

The results of all the fits to data are shown in~\cref{fig:sea_scaling}. 
The \dquotes{M. Polished (2020)} fit was performed on datasets from Runs 4, 6, 7, and 9, in which several electrode separations were tested (and therefore SEA values), using a pair of mechanically polished electrodes.
Runs 5 and 8 were excluded from the fit: the former because of an overpressure event during condensing and the latter due to the use of a different set of electrodes with which significant different results were observed. 
Run 9 was performed with the same set of electrodes as Runs 4--7, but they were re-polished because it was observed that the previous experimental runs had resulted in pitting of the electrodes (see \cref{sec:observations}).
Additionally, a fit was performed on the datasets corresponding to the passivated electrodes (Run 10), which is labelled as \dquotes{Passivated}.
These electrodes, which were passivated with citric acid, were observed to outperform the mechanically polished electrodes, especially at smaller SEA.
However, we note that this fit is based on three data points and a finer sampling of SEA, as well as an extension to higher values, is needed to verify this trend. Overall, a downwards trend in breakdown field with stressed area was observed in all cases.

The best-fit parameters to the power law in \cref{eq:weibull_area_dependence}, as applied to $E_2$, along with their Pearson correlation coefficient, are tabulated in \cref{tab:powerlaw_fits}.
The error bars correspond to the quadratic sum of the statistical uncertainty, taken from the covariance matrix of the Weibull fit parameters, and a systematic uncertainty propagated from the systematic uncertainty in the measurement of the electrode separation (\SI{0.03}{mm}).
The errors in SEA are obtained via the FEniCS simulation after accounting for the cathode tilt angle at each run (see Section~\ref{sec:E_sims}). 
The summed squares of residuals along both axes were minimized to obtain these fits.


One of the current generation liquid xenon TPCs is the LUX-ZEPLIN (LZ) experiment. 
The LZ cathode ring has a simulated stressed area of approximately 940 cm$^2$.
Evaluating the model from this work for the mechanically polished electrodes gives a predicted shifted scaled parameter ($E_2$) of 28.2 $\pm$ 0.5 kV/cm.
This is a revision downwards from the prediction in Ref.~\cite{tvrznikova_direct_2019}, shown in blue in the top figure of~\cref{fig:sea_scaling}. 
Using the model from XeBrA's 2018 analysis, the predicted breakdown field at the LZ cathode ring SEA is $70.4 \pm 3$ kV/cm.
We believe the discrepancy results mainly from the following changes: the uncertainty on the cathode tilt angle was taken into account (instead of assuming a 0$^{\circ}$ tilt of the cathode), each run consisted of several days of operation in which the xenon was continuously being circulated (as opposed to one very long day of data-taking), and a new method to mitigate bubble formation in the bulk was introduced. 
To provide some reference, the maximum design voltage in LZ was set at 50 kV/cm~\cite{Mount:2017qzi}, which is intermediate between these two extrapolated values.
In addition, the predicted breakdown voltage for a 40-tonne active volume TPC (e.g.~DARWIN~\cite{Aalbers:2016jon}), assuming that SEA scales with mass to the power of $2/3$, would be $20.7 \pm 0.4$ kV/cm.
This is illustrated in the bottom panel of \cref{fig:sea_scaling}, where the estimated stressed area of the cathode ring for LZ and two hypothetical Generation-3 (G3) experiments are shown.
Consequently, careful engineering of the electrostatics of any future, large TPC will be critical in achieving the desired fields. 
This analysis indicates that future LXe experiments of the scale of DARWIN should design around a maximum electric field of \SI{20}{kV/cm} on their cathode rings in order to have some safety margin. However, we note that improvements on cathode design could raise this threshold.

\begin{table}[]
    \centering
    \begin{tabular}{|c|l|l|l|}
    \hline
         Model & C [kV/cm] & b & $\rho_{Cb}$\\
          \hline 
         M. Polished (2020)& 169.5 $\pm$ 3 & 0.262 $\pm$ 0.006 & -0.995\\
         Passivated (2020) & 530 $\pm$ 28 & 0.667 $\pm$ 0.02 & -0.920\\
         M. Polished (2018) & 171.5 $\pm$ 8& 0.13$ \pm$ 0.02 & -0.995\\
         \hline
    \end{tabular}
    \caption{XeBrA power law fit parameters to \cref{eq:weibull_area_dependence}. 
    $C$ is a multiplicative constant, and $-b$ is the coefficient of the SEA in the power-law scaling. 
    }
    \label{tab:powerlaw_fits}
\end{table}

Lastly, and while not central to this analysis, the theory that field emission from local asperities initiates dielectric breakdown remains a possible interpretation of the data.
Without a high precision direct measurement of DC current, the area of the field emitter can not be unambiguously determined.
Following the convention in Ref.~\cite{Phan:2020nhz}, we make the assumption that the cumulative hazard is proportional to the FN field emission current.
In such case, the slope of the FN plot ($\log(H(E)/E^2)$ vs $1/E$) combined with the work function of stainless steel provides a measurement of $\beta$, the local field enhancement factor.

\begin{figure}[tbp]
    \centering
    \includegraphics[width=0.9\columnwidth]{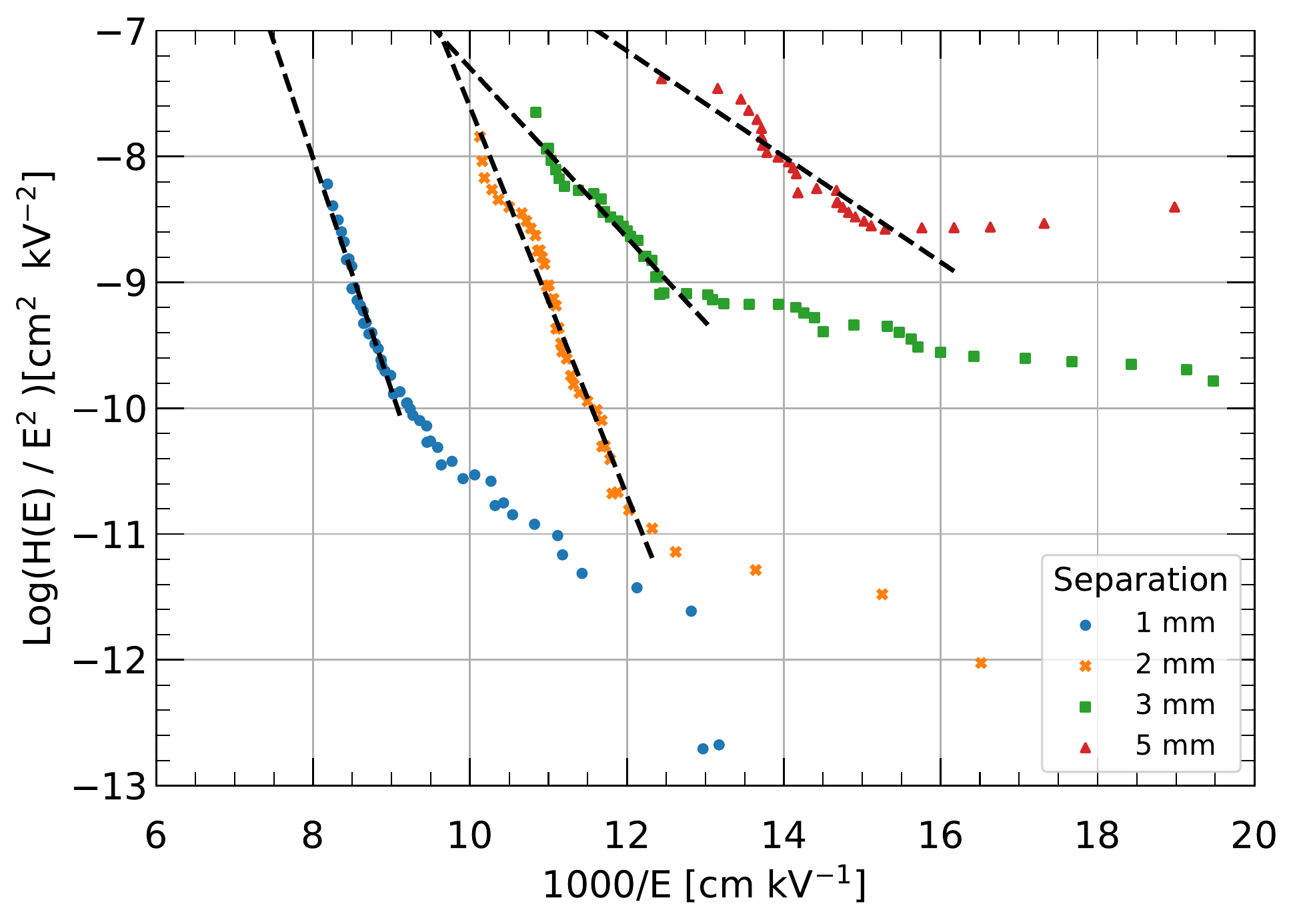}
    \caption{Fowler Nordheim plot for Run 7. 
    $H(E)$ represents the cumulative hazard due to the electric field.
    All datasets shown here have a pressure setpoint of 1 bar absolute and a ramp speed of \SI{100}{V/s}.
    An increase in hazard, identified as a pseudo-current ($J$), occurs as the separation distance increases. Black dashed lines represent a power-law fit to the linear regions at high field. 
    }
    \label{fig:fnplot}
\end{figure}

An example of such plot for Run 7 (using polished electrodes) is provided in \cref{fig:fnplot}.
Deviations from linearity are observed at lower fields and are not included in the final fit.
A linear regression performed on the quasi-linear regions of the highest fields, yields $\beta$ values ranging from $100$ to $2000$ for the different electrode separations that were scanned. Such disparity among the estimated $\beta$ values implies the possibility of emitter locations changing over the course of the run. Similar results were observed for different runs. Albeit not impossible, this is an unlikely result.
Consequently, the FN model was not utilized in the final analysis, but field emission from the cathode remains a possible explanation for a starting mechanism of dielectric breakdowns in the liquid.


\subsection{Pressure}
\label{sec:Pressure} 

Liquid pressure was scanned in addition to separation distance. Pressure values ranging from 1.5 to 2.2 bar absolute were scanned across all runs at a fixed ramp rate of \SI{100}{V/s} (see the complete list of configurations in \cref{tab:run_fits}).
The separation distance was fixed at \SI{2}{mm} and the pressure was adjusted in steps.
No unambiguous relationship between the system pressure and breakdown voltage was observed for any of the runs.
An approximately $10$\% variation in breakdown field between datasets at a different pressure was observed, however this result was insensitive to the direction of change. 
During one of the runs (Run 8) the pressure was increased monotonically and the breakdown field was observed to be more consistent, yet still largely insensitive to pressure changes. 

We suspect that the overall thermodynamics of the system have a larger impact on the breakdown field than anticipated. Pressure has a direct dependence on the temperature and circulation flow rate in the system, making it difficult to control independently. 
Changes to the operational procedure of XeBrA might help to control for some of these confounding variables better in the future.

\subsection{Ramp speed}
\label{sec:Speed}

If breakdown risk were modeled as a rate per unit time, it would be expected that a faster ramp speed would lead to larger breakdown fields.
As \cref{eqn:hazard_time} shows, breakdowns cannot generally be described as a simple threshold.
By contrast, there is some stochastic risk of failure per unit time, and increasing the high voltage faster entails spending less time at any particular field.
We caution that the experimental framework here does not describe the operating conditions of typical TPCs, where the voltage is ramped to a specified voltage and held there for potentially years.
Instead, what \cref{eqn:hazard_time} does is to disambiguate the variance of the breakdown values (i.e.~the Weibull shape parameter) from the effect of the \dquotes{history} of each ramp.
An example of this would be the accumulation of damage to the electrodes or the medium from previous discharges.

In XeBrA, a larger breakdown field was indeed observed with increasing ramp speed with passivated electrodes, but inconclusive results were found with the non-passivated, polished electrodes.
In the range of 100 to 150 V/s, both sets of electrodes show an increase in breakdown field of comparable magnitude. 
At \SI{200}{V/s} the positive trend continues for the passivated electrodes but reverses for the polished electrodes. 

Hence, no definitive model of breakdown fields as a function of ramp speed was found.
The observed variation between datasets with different ramp speeds is approximately 10\%.
Future runs of XeBrA will explore more values of the ramp speed scan than the ones presented in this work.


%

\subsection{Precursors}
\label{sec:precursors}

It is of particular interest to determine whether an imminent breakdown can be sensed, such that possible interventions may be taken to avoid damage to the detector.
If breakdowns are initiated by field emission near local asperities, one might expect an increase in current into the anode as the cathode voltage increases, as both the FN current and any multiplication effects increase in magnitude.
In XeBrA, precursor discharges were typically observed before breakdowns. 
These appeared to be similar to breakdowns in that the current rapidly pulses, but the amplitude of these pulses were orders of magnitude lower.

\begin{figure*}[tbp]
\centering
\includegraphics[scale = .4]{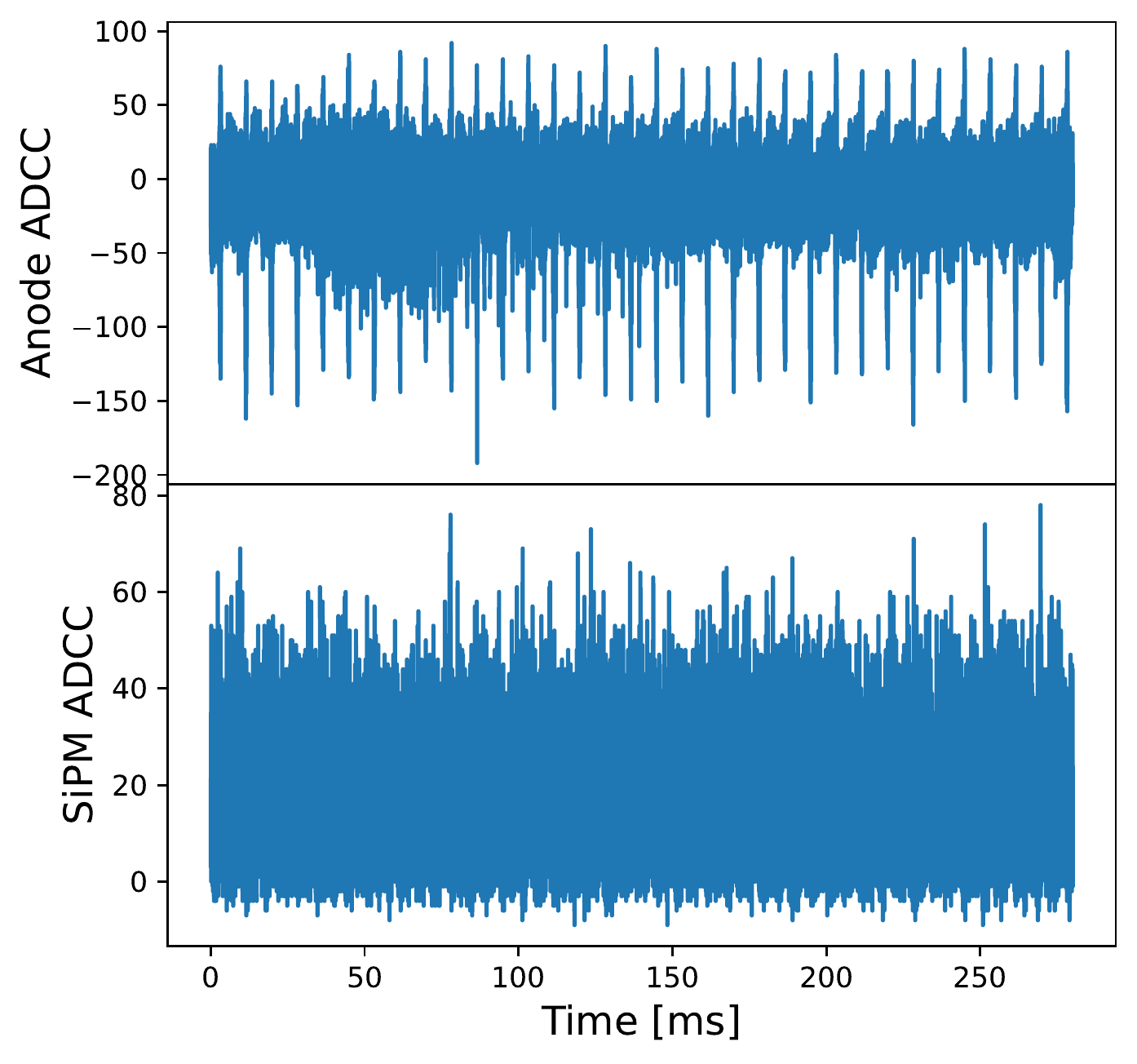}\qquad
\includegraphics[scale = .4]{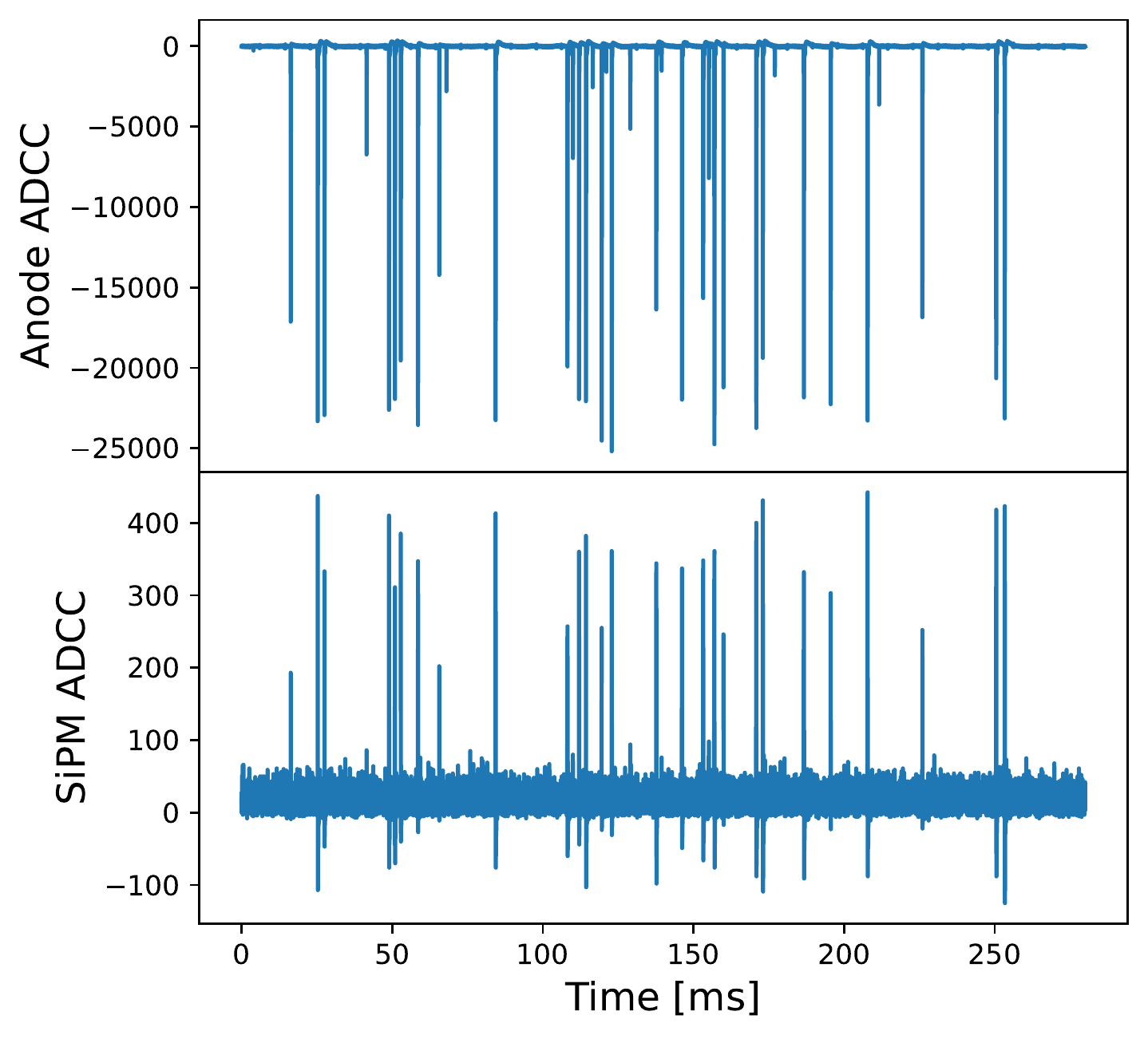} \qquad
\includegraphics[scale = .4]{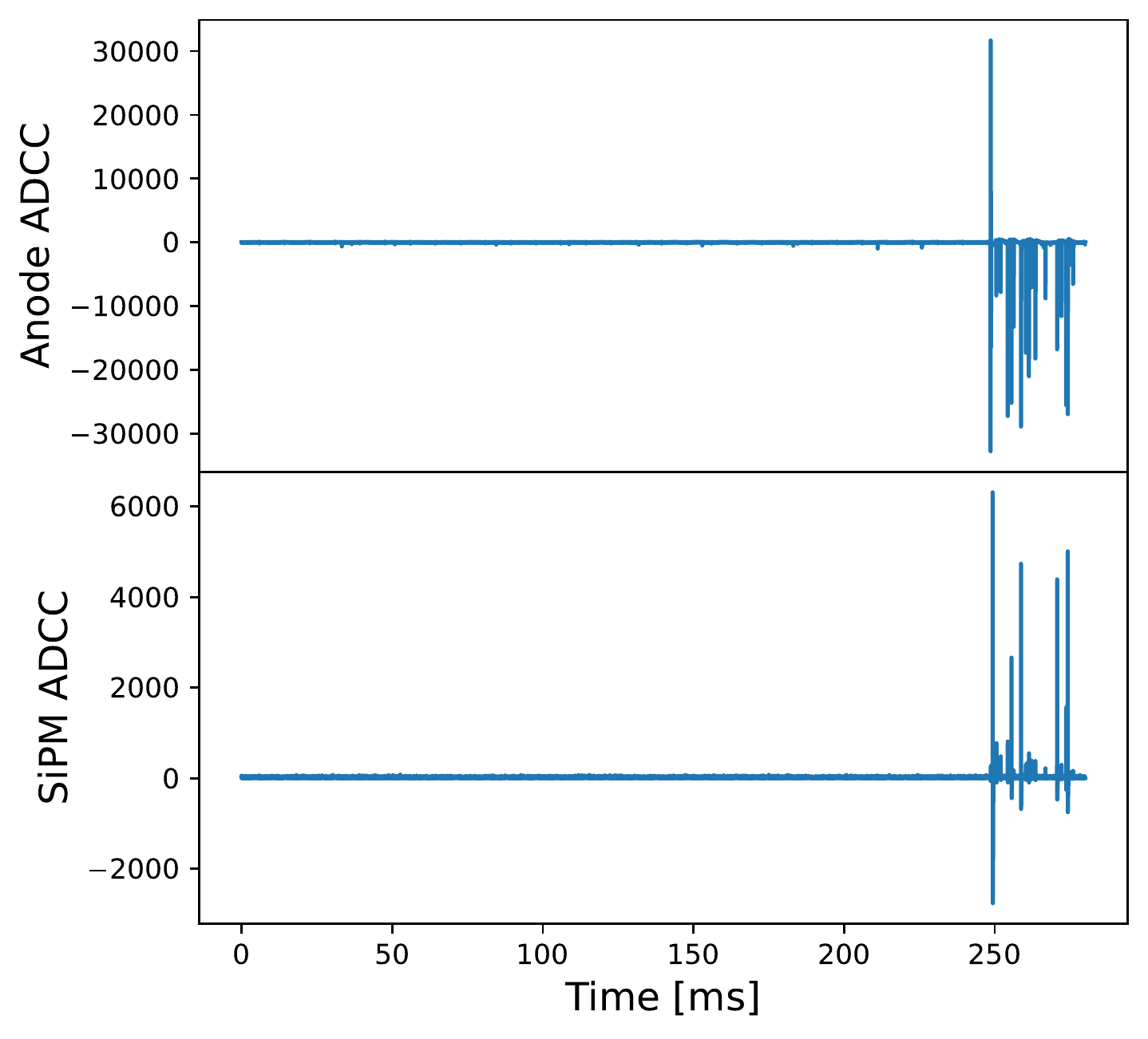}\qquad
\includegraphics[scale = .4]{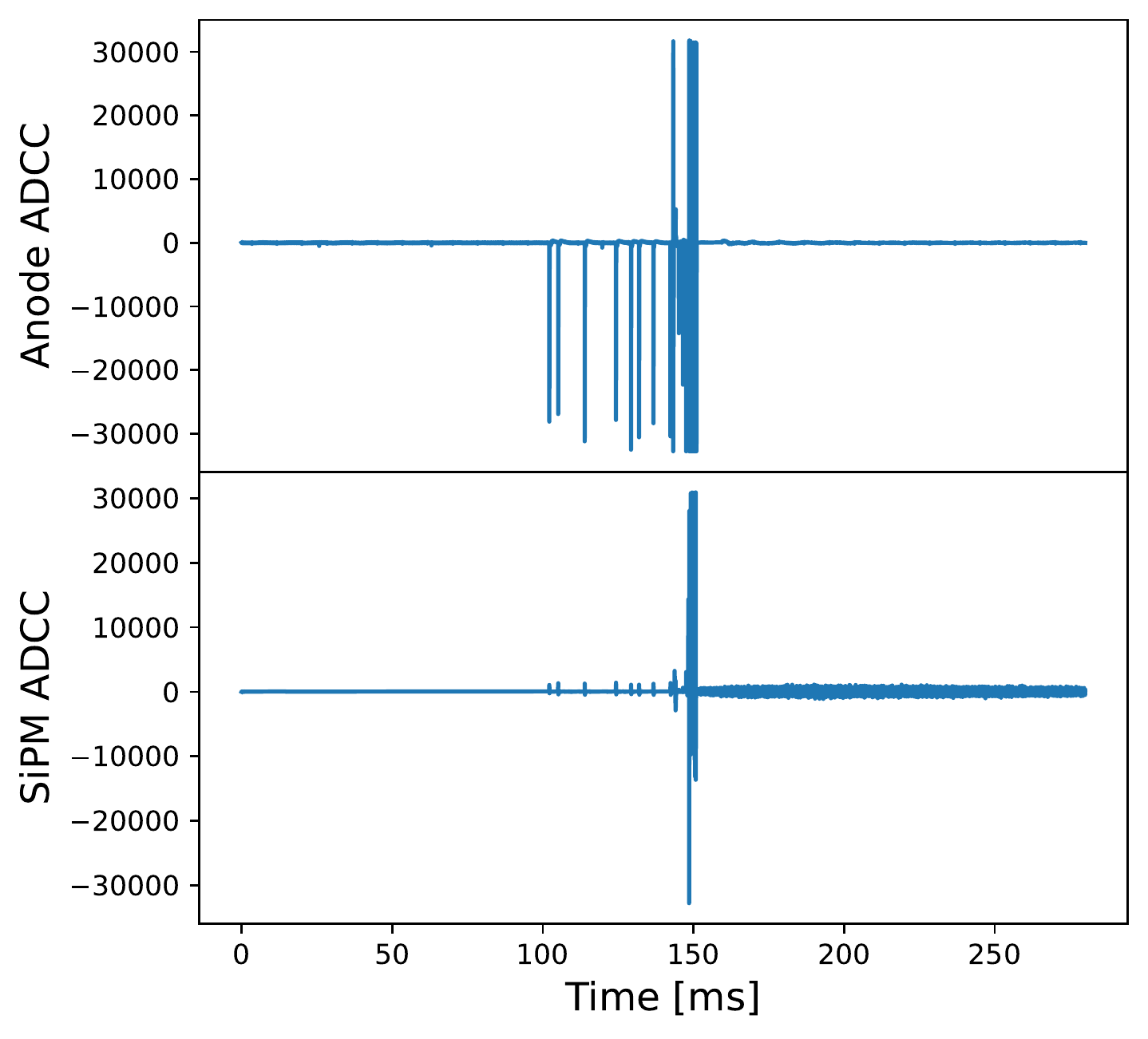}

 \caption{Various categories of event acquisitions.
  \textit{Top left}: An example of background noise. 
The oscillatory behaviour is removed through a filtering technique.
 \textit{Top right}: an example of the coincidence between anode and SiPM pulses. This event was recorded close to the breakdown point.
 \textit{Bottom left}: A subthreshold discharge. These occasionally occur during a ramp, but did not trigger an HVPS fault signal.
 \textit{Bottom right}: A true breakdown event, coincident with an HVPS fault signal.
    }
    \label{fig:Waveforms}
\end{figure*}

The digitized signals from the anode current and SiPM were analyzed offline in order to locate short pulses.
The waveforms were processed in blocks of 10$^{6}$ samples and filtered using a zero-sum difference-of-Gaussians filter.
Peaks in the filtered response are treated as a pulse, with boundaries at the zero crossings on either side of the peak.
The particular widths of the filter were tuned to the particular noise and electronics response of the run being analyzed.
A semi-regular chirp-like signal was present in background acquisitions (see \cref{fig:Waveforms}, top left), which was filtered out with a separate algorithm.
The signal is first filtered to find the standard deviation, calculated over a fixed-length moving window.
Regions where the ratio of the signal amplitude to the standard deviation were sufficiently low were replaced with a low-pass filter value.
To further filter the chirp-like pulses, the condition FWHM < \SI{8.4}{\us} was applied. Also, only anode pulses of negative polarity were selected.
Several quantities were calculated on each pulse and the slow control logs provided information on the detector conditions during each pulse.

An algorithm to locate breakdowns in the DAQ was used to terminate each ramp and group the pulses into an object for analysis. 
During a breakdown, a large overshoot of the anode current signal was frequently seen, leading to what appear to be large surges of positive current into the anode.
These also occasionally occurred in the seconds preceding breakdowns. Three such surges in one event window were required to flag a breakdown.
An example of a sub-threshold discharge and a true breakdown event are shown in the bottom left and bottom right of \cref{fig:Waveforms}, respectively. 

The charge amplifier signal frequently saturated the dynamic range of the DAQ, and large anode pulses were almost always associated with a corresponding pulse in the SiPM channel, as shown in the last three plots in \cref{fig:Waveforms}.
Typically, an increase in anode pulse rate of around a factor of 100 was observed in the preceding 60 seconds to a breakdown.
Due to the lower light collection and high dark rate compared to the PMT, it was not possible to observe a large amount of light in the SiPM until just a few \SI{}{ms} before a breakdown.
The anode pulses have a distinctive spectrum of heights. 
Most pulses were barely above threshold, but the overall distribution was bimodal, which was particularly evident when the gain was lowered on the charge amplifier.


%
\begin{figure}[tbp]
    \centering
    \includegraphics[width=0.95\columnwidth]{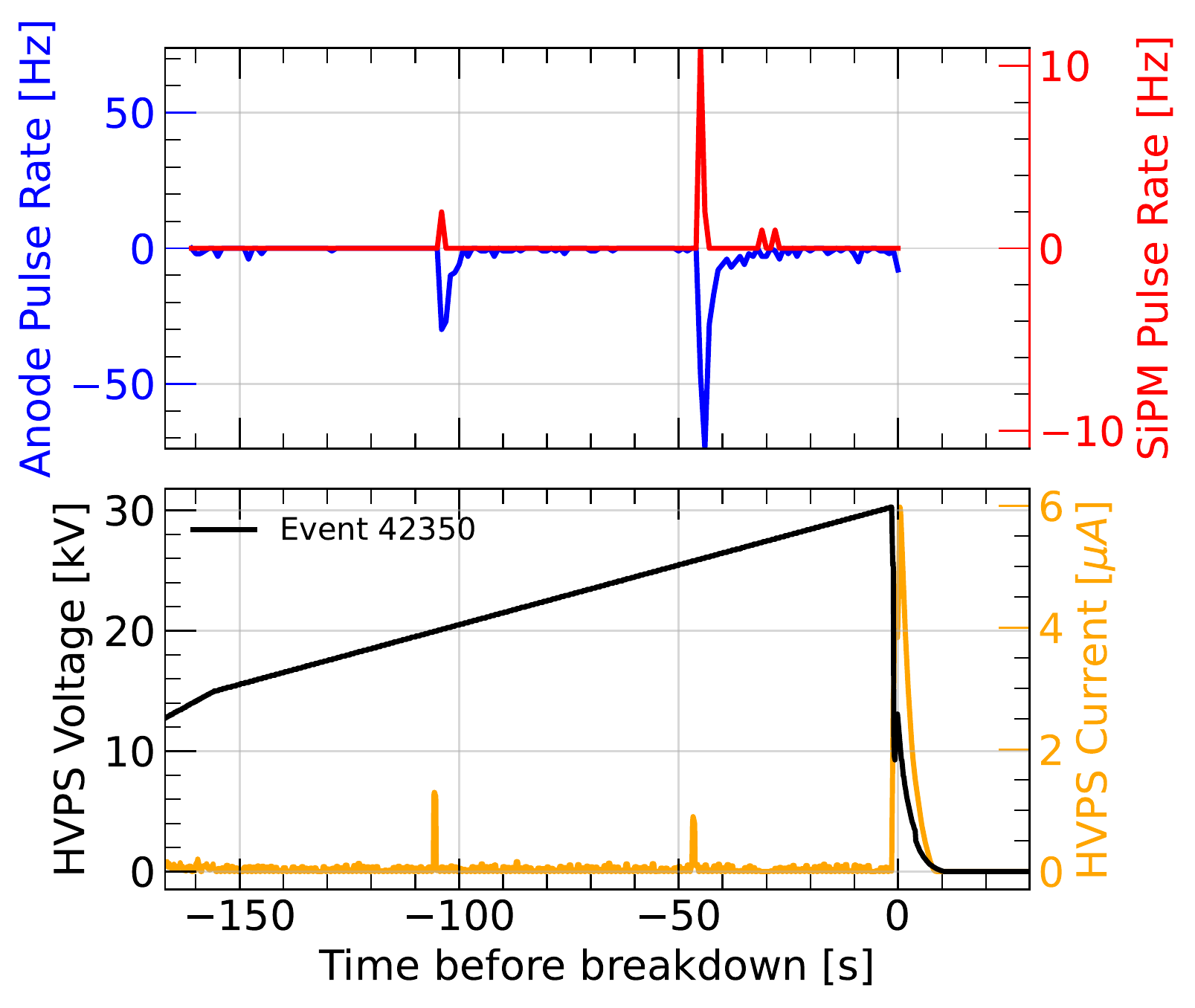}
    \caption{
    \emph{Top:} Pulses found over the course of a single ramp.
    The anode and SiPM pulses are nearly coincident with one another. 
    The anode pulses have a better signal-to-noise ratio, and therefore smaller peaks can be resolved on this scale.
    \emph{Bottom:} Complementary information obtained from the slow control logs for the same ramp. 
    The burst in current at time 0 is not the breakdown itself, but rather the discharge of current through the HV power supply due to the sudden drop of electric potential energy.
    }
    \label{fig:breakdown}
\end{figure}
\begin{figure}[tbp]
    \centering
    \includegraphics[width=0.95\columnwidth]{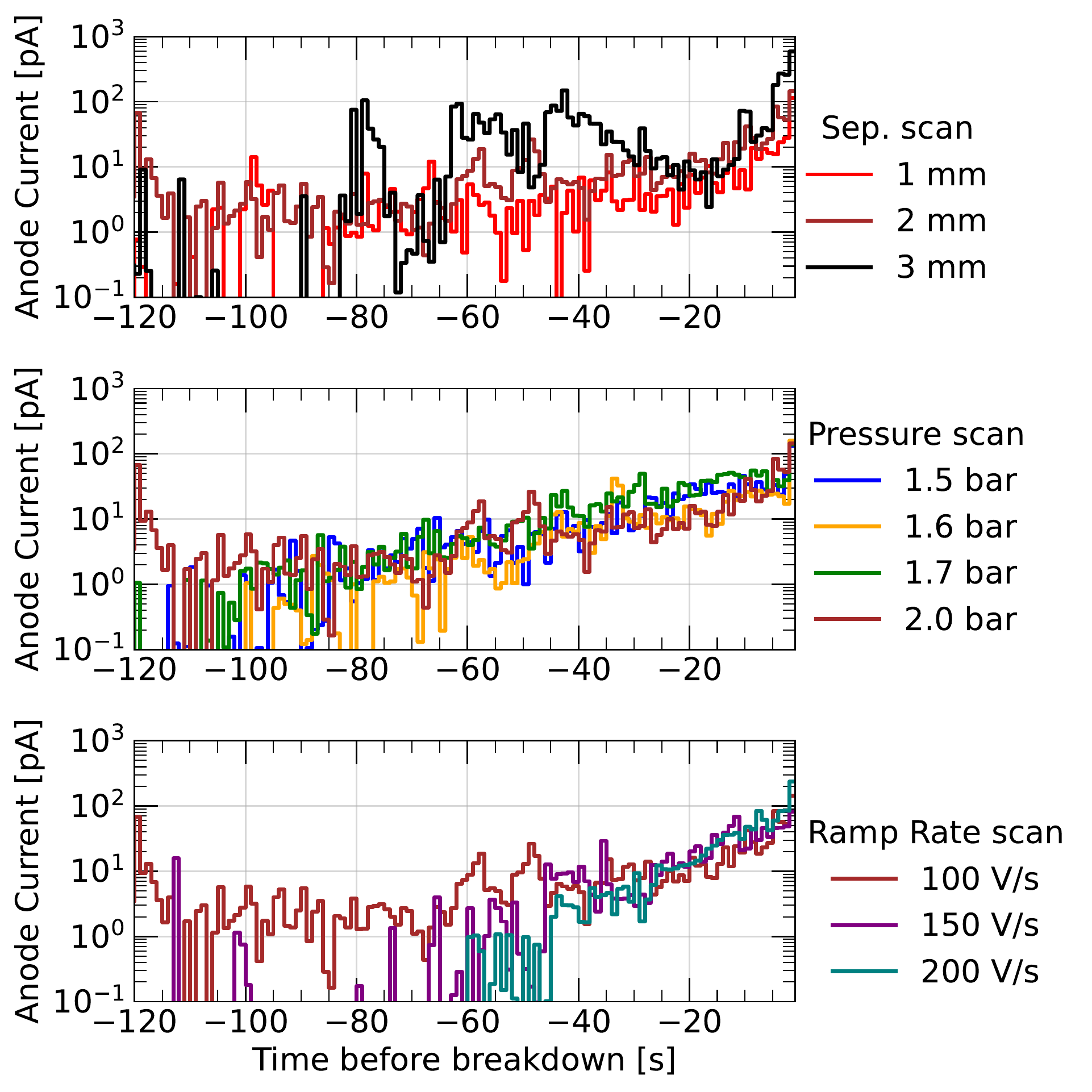}
    \caption{
    Distributions of anode current preceding a breakdown for the following parameter scans in Run 10: electrode separation (top), pressure (middle), and ramp rate (bottom). The anode current is averaged over in time bins of 1 second. Pulses within 2 event windows (approximately 750 ms) of a subthreshold breakdown (defined in text) were discarded. A general rise in anode current in the few tens of seconds before breakdown is observed in all cases.
    }
    \label{fig:Run9_chargetime}
\end{figure}
The anode current pulse rate was generally higher in runs with larger SEA, while pressure and ramp speed had no significant effect on the time profile of precursor pulses.
The precursor rate within any particular ramp did not rise continuously, but instead appeared to spike in rate several times preceding a breakdown (see an example in \cref{fig:breakdown}). Overall, a consistent rise in anode current before breakdown was observed. For instance, \cref{fig:Run9_chargetime} shows the average current (as measured by the charge amplifier) over time bins of \SI{1}{s} for the different parameter scans in Run 10.
These plots show a first surge in current around 50--\SI{60}{s} and a rapid increase in current in the last few seconds.

\begin{figure}[tbp]
    \centering
    \includegraphics[width=0.9\columnwidth]{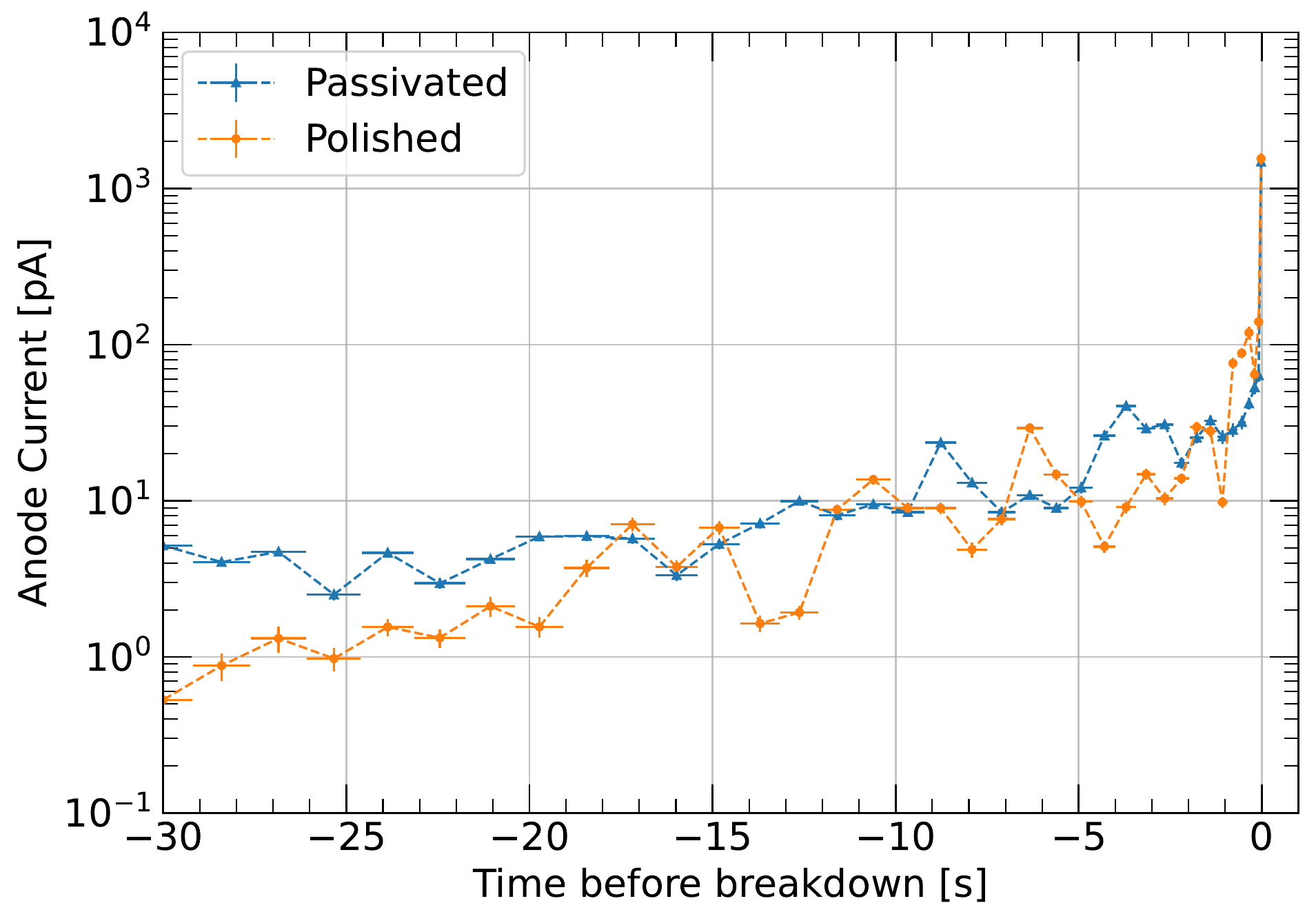} \qquad
    \includegraphics[width=0.9\columnwidth]{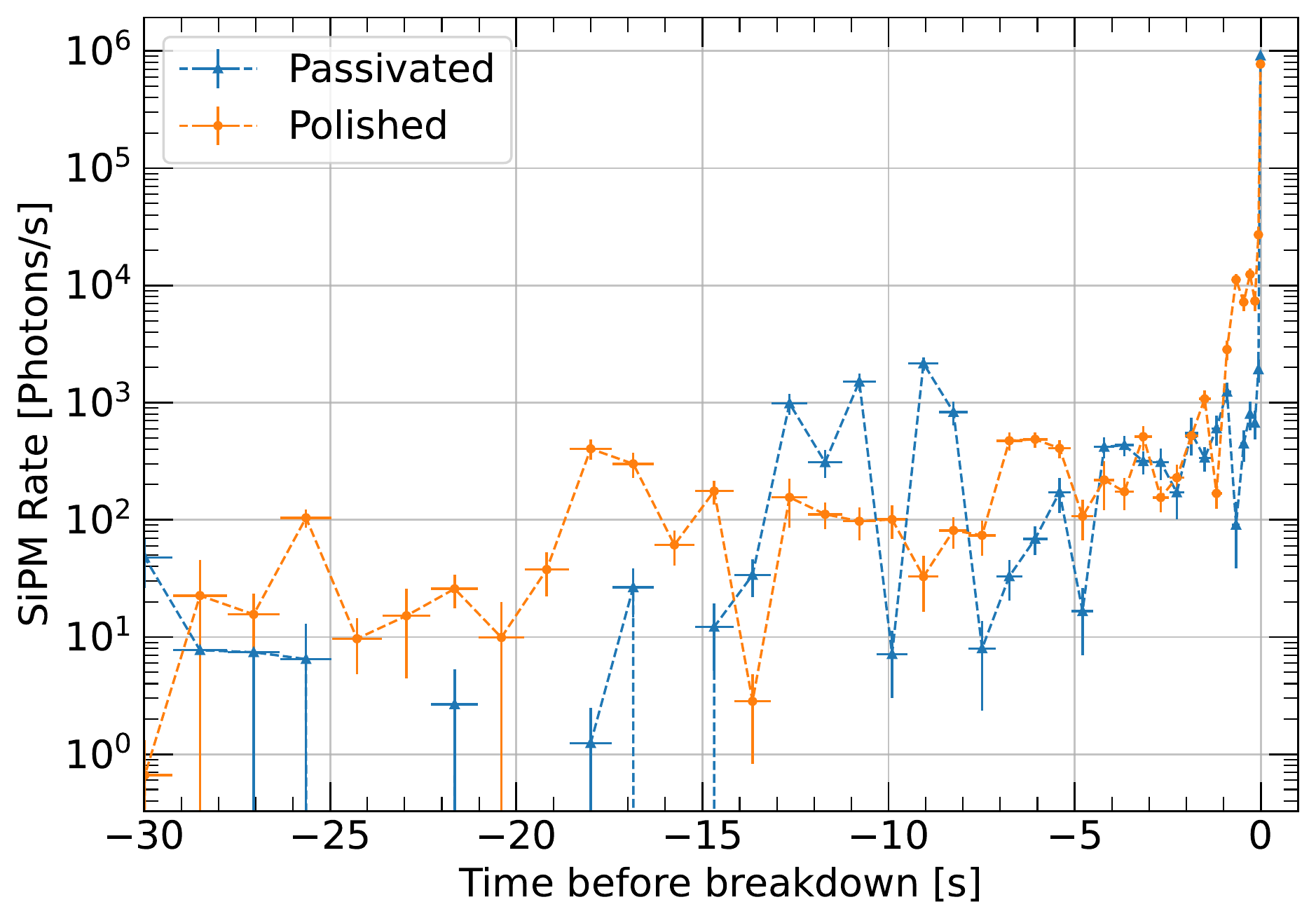}
    \caption{Comparison of anode current (top) and SiPM rate (bottom) in the 30 seconds preceding a breakdown between mechanically polished electrodes (Runs 9, orange line) and passivated electrodes (Run 10, blue line). The anode current is averaged over in time bins of \SI{1}{s}. An electrode separation of \SI{2}{mm} is considered in both cases. The two sets of electrodes perform similarly, except for the last few seconds in which the polished electrodes exhibits increased activity relative to the passivated ones.
    }
    \label{fig:ElectrodeComparison}
\end{figure}

Comparing the polished and the passivated electrodes, we observed that passivation only has a small effect on the preceding activity before a breakdown.
As \cref{fig:ElectrodeComparison} shows, there is only a moderate suppression in both the anode current and SiPM rate in the immediate few seconds before a breakdowns when comparing Run 9 (mechanically polished electrodes) to Run 10 (passivated electrodes).

\begin{figure}[tbp]
    \centering
    \includegraphics[width=0.9\columnwidth]{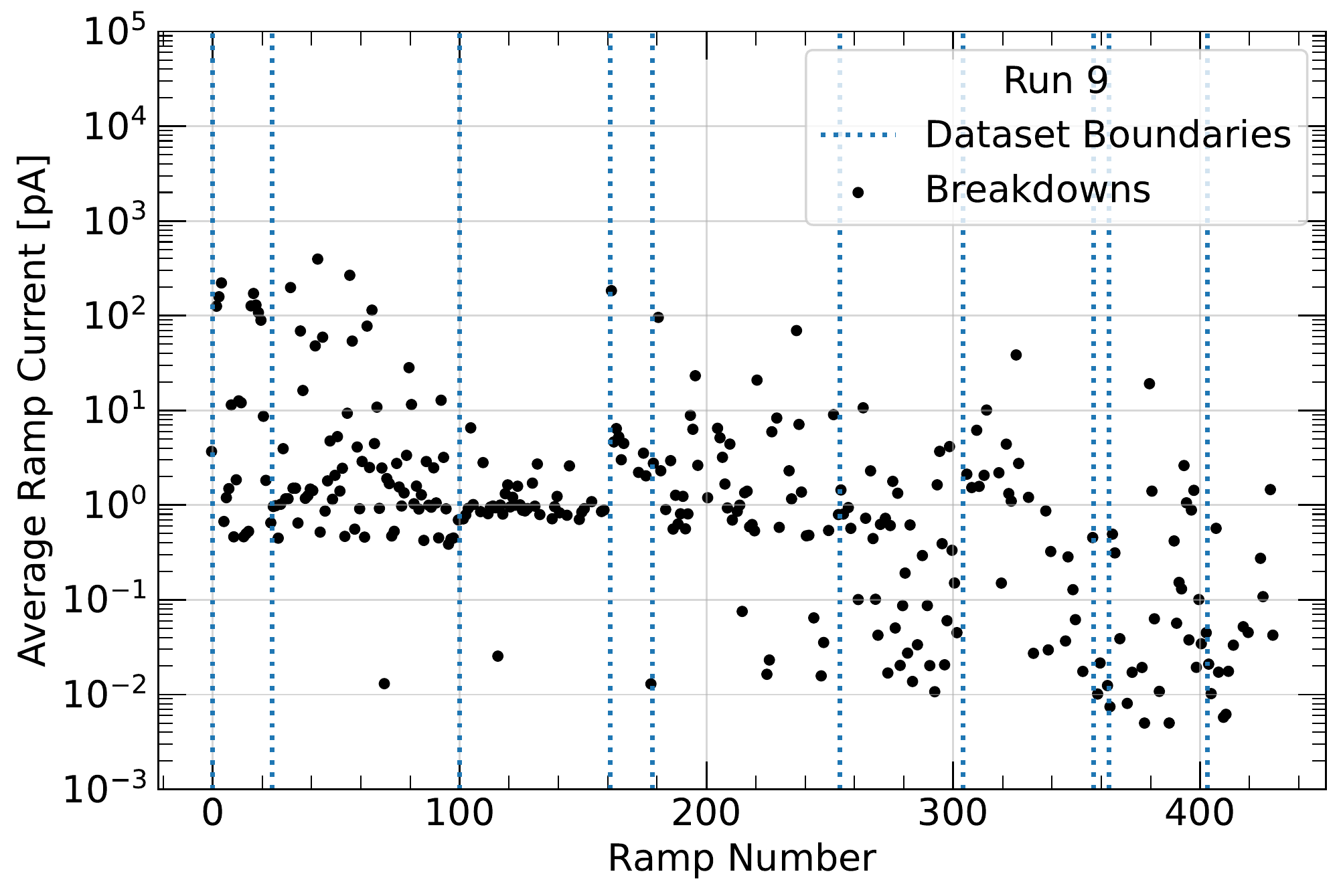} \qquad
    \includegraphics[width=0.9\columnwidth]{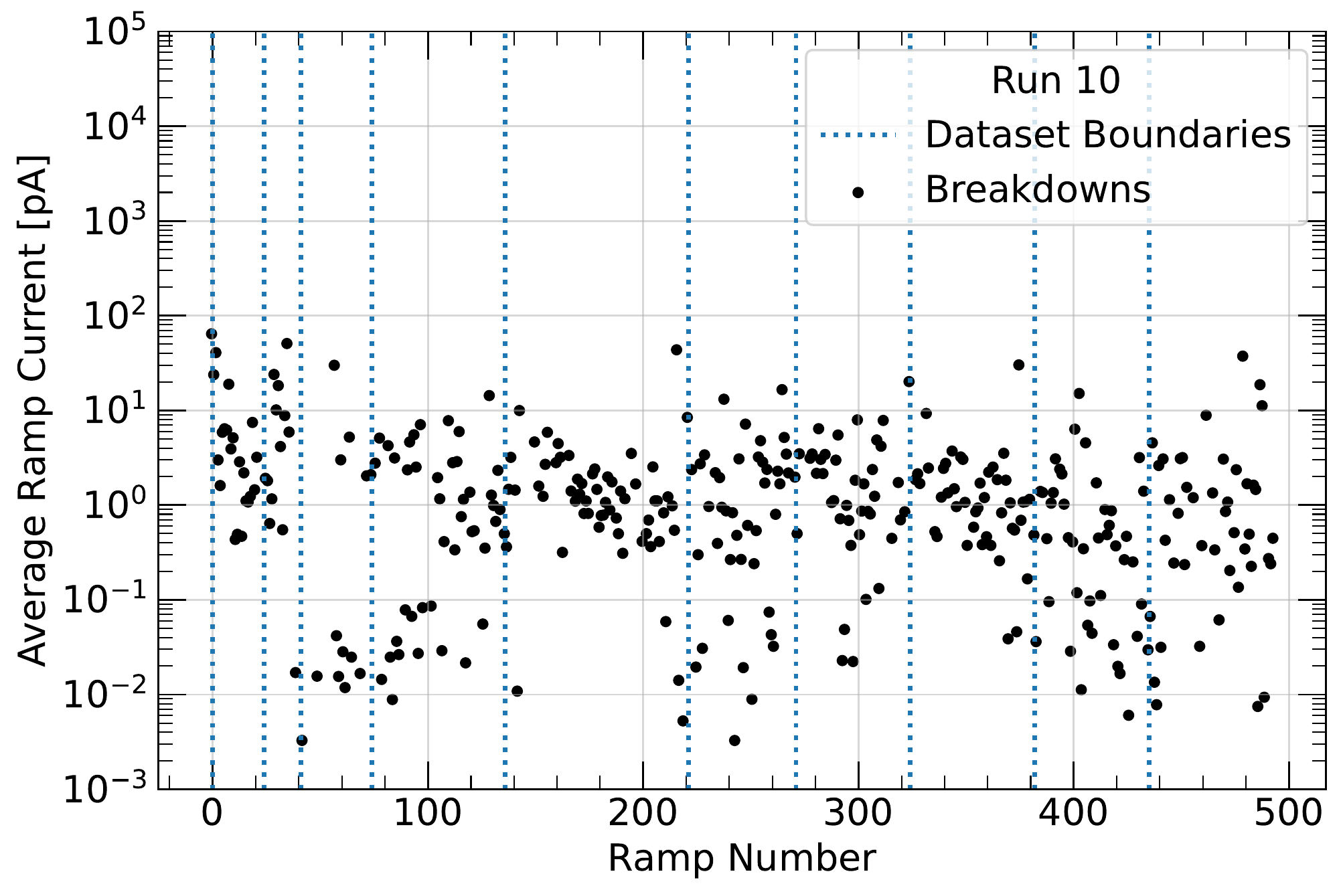} \qquad
    \caption{
    Anode current as a function of HV ramp number for the entire Run 9 (top) and Run 10 (bottom), in which a fresh set of polished and passivated electrodes were used, respectively. The anode current is integrated between the first 45 and 120 seconds from the start of the HV ramp, with a minimum gap of 20 seconds between the start and end of each ramp. A downward trend within each run is observed in both cases, which can be attributed to conditioning of the electrodes. 
    }
    \label{fig:Conditioning_plot}
\end{figure}

A conditioning effect over the course of a run was also observed. This is illustrated in \cref{fig:Conditioning_plot} for Runs 9 and 10.
The current into the anode was averaged over a time period well separated from either the breakdown itself or the initial ramp.
The region of \SI{45}{s} to \SI{120}{s} from the start of the ramp was chosen to obtain a large window and avoid capturing the immediate few seconds before a breakdown.
A downward trend is observed in both cases, indicating a reduction in charge being emitted as more ramps were conducted during each run.





\subsection{Position reconstruction}
\label{sec:pos_recons}

Videos were recorded during all runs in order to reconstruct the position of breakdowns. Unfortunately, the number of videos acquired in Runs 5 through 8 was insufficient and some Run 9 videos were compromised by the field of view of one of the cameras being blocked.
From the remaining videos, the positions of breakdowns were reconstructed: Run 9 at 1 mm and Run 10 data at 1, 2 and 3 mm electrode separation distances.
The distributions of these breakdowns were compared to the FN model of field emissions using simulations of the electric field for the XeBrA geometry.
For this comparison, the tilt between the electrodes determined from the tilt estimation procedure, found in Table ~\ref{tab:tilts}, was used when simulating the electric field for a given Run. 
The resulting electric field distribution was then used to find the corresponding FN surface current density.

\begin{figure}[tbp]
    \centering
    \includegraphics[width=0.9\columnwidth]{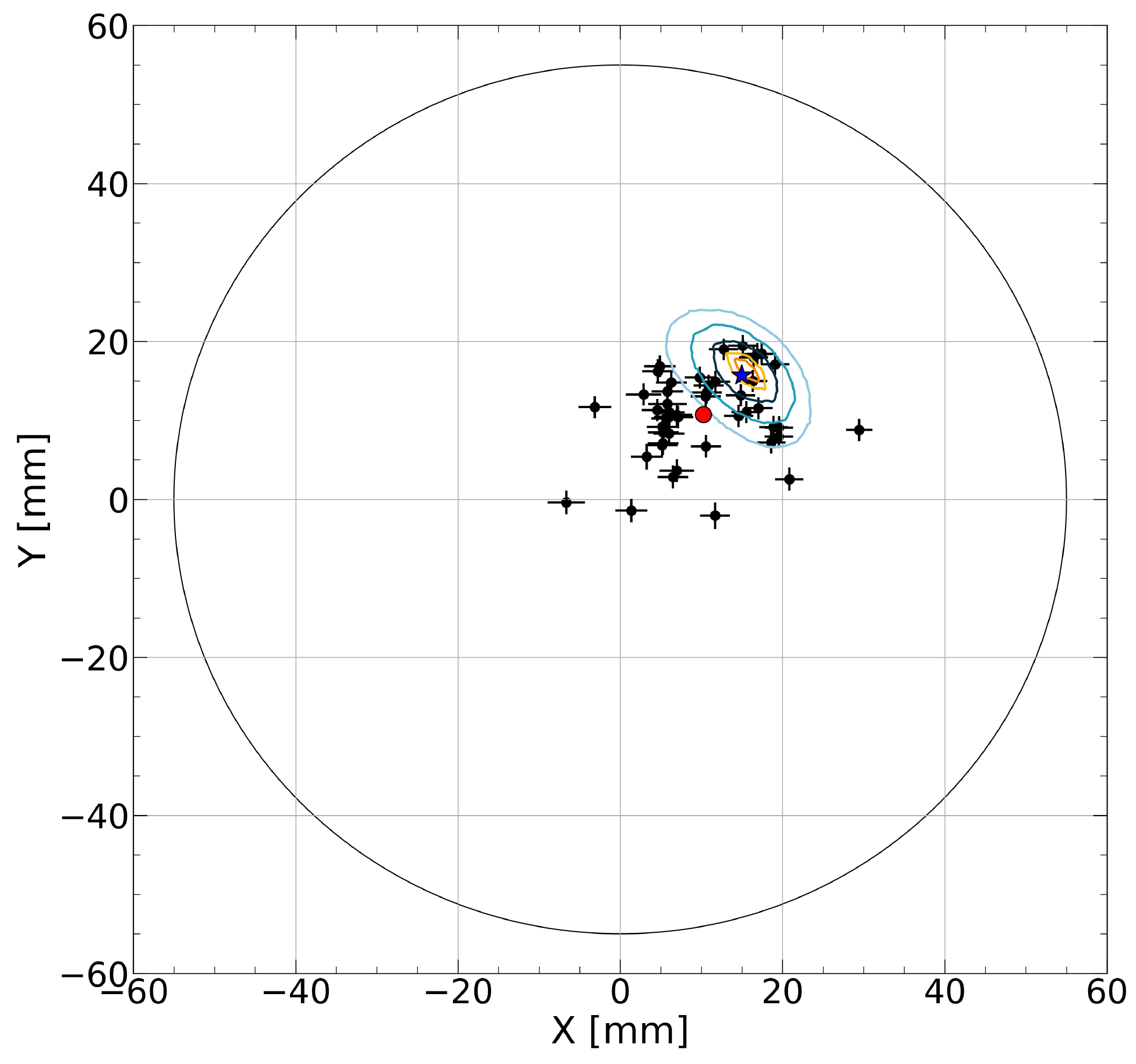}
    \caption{Comparison between breakdown distributions and FN current distributions from simulations for Run 9, 1 mm data. The breakdown distribution (points) are overlaid with the FN current densities (isocontours). Isocontours correspond to the 90, 80, 60, 30, and 10 $\%$ of the maximum FN current. The centroids of each distribution are marked with a red dot for data and a blue star for simulation, respectively.}
    \label{fig:posreco}
\end{figure}

For each breakdown distribution, the Gaussian Kernel Density Estimate (KDE) was found using the scikit-learn library~\cite{scikit-learn}.
The centroids of the KDE and maximum values of the FN surface current were recorded and are given in Table ~\ref{tab:centroids}.
The FN surface currents were rotated to minimize the distance between the centroid of the KDE and the maximum of the FN surface current.
As an example, the subsequent FN surface current densities, along with the breakdown distributions, are shown in Fig.~\ref{fig:posreco} for Run 9, 1 mm data. 

Overall, the position distribution of breakdowns shows a direct relationship with the simulated SEA.
More specifically, broader SEA lead to broader breakdown distributions. 
When these distributions are compared to the FN surface current densities, related to probability of electron emission, the distances between the centroids of the data and maximums of the FN currents are, on average, found to be around 9 mm (see \cref{tab:centroids}).
These results are favourable towards the FN model, but more detailed comparisons between reconstructed positions and simulations are needed, which will be the object of study of future XeBrA runs. 


\begin{table}[hb]
    \centering
    \begin{tabular}{|c|l|l|l|l|l|}
    \hline
    Run & Separation [mm] & N$_{\text{rec}}$ & Centroid & Maximum & Distance [mm]\\
    & & & of data & of FN & \\
    \cline{4-5}
     & & & x, y [mm] & x, y [mm] & \\
    \hline
         9  & 1 & 47 & 10.2, 10.8 & 15.0, 15.7 & 6.9\\
         10 & 1 & 46 & 3.9, 3.0 & 12.4, 9.7 & 10.8\\
         10 & 2 & 50 & 5.0, 0.4 & 15.7, 1.3 & 10.8\\
         10 & 3 & 50 & 9.1, 4.8 & 14.0, 7.4 & 5.6\\
         \hline
    \end{tabular}
    \caption{Centroids of both data and simulations for those datasets with the largest number of reconstructed positions (N$_{\text{rec}}$). The last column shows the distance between the centroid of the reconstructed positions and the maximum of the electric field simulation.}
    \label{tab:centroids}
\end{table}

\subsection{Visual observations}
\label{sec:observations}

The presence of viewports oriented at right angles to the electrodes facilitated visual observations of a variety of phenomena.
Critically, this provided certainty that HVPS current trips corresponded to breakdowns between the electrodes as well as verifying the bubble-free state of the liquid inside the chamber. 

\begin{figure}[tbp]
   \centering
    \includegraphics[trim=0 6cm 0 6cm, clip, width=0.5\columnwidth]{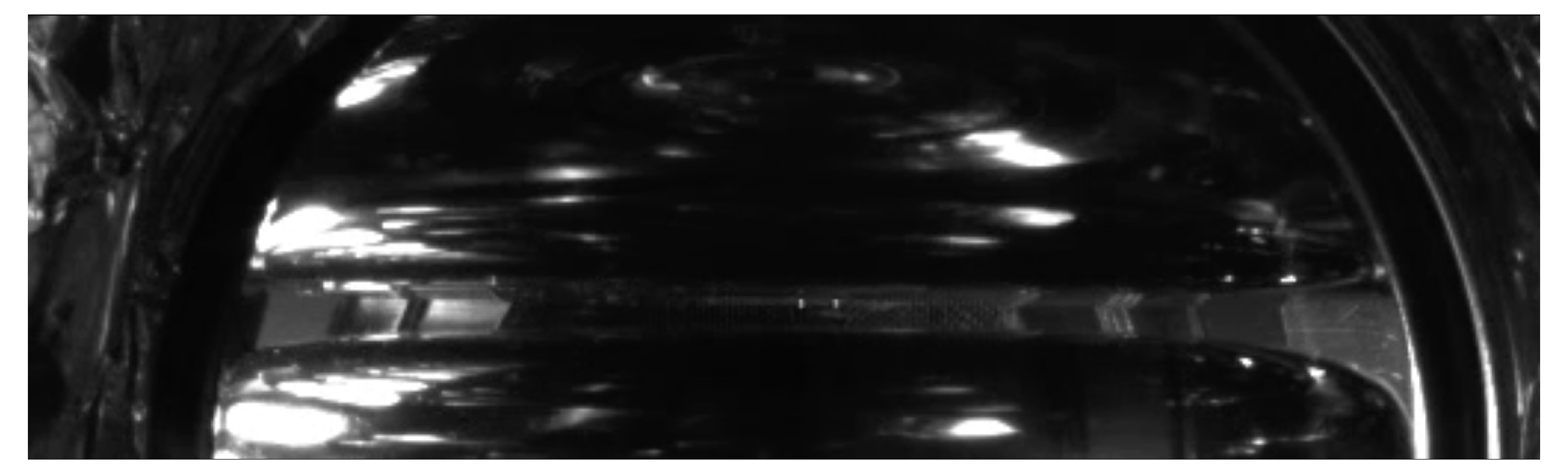}
    
    \includegraphics[trim=0 6cm 0 6cm, clip, width=0.5\columnwidth]{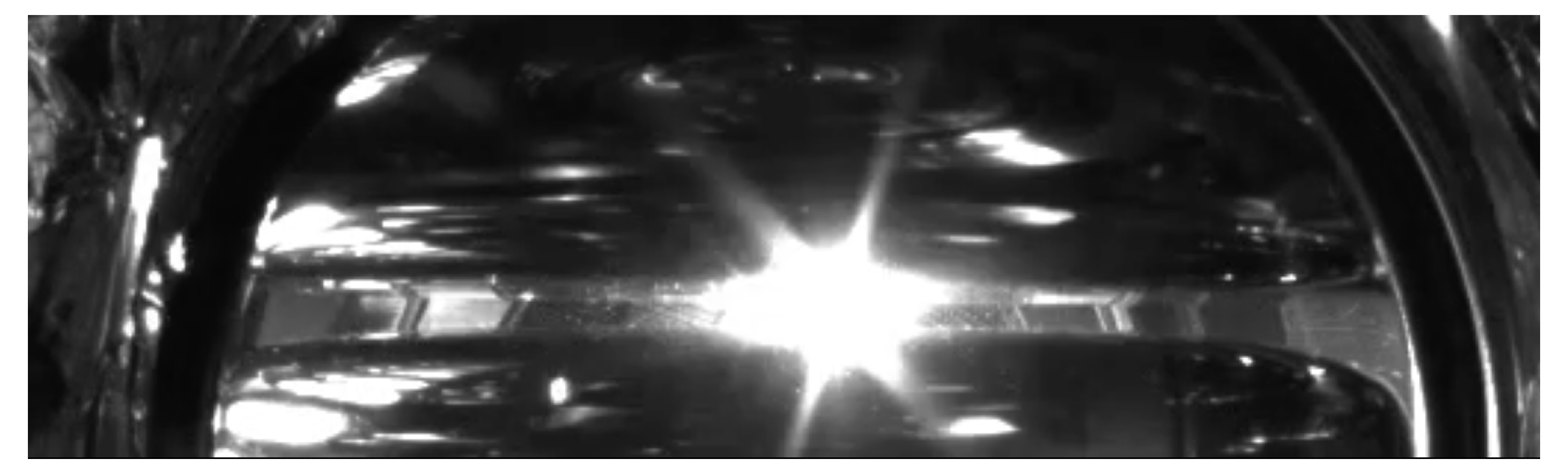}
    
    \includegraphics[trim=0 6cm 0 6cm, clip, width=0.5\columnwidth]{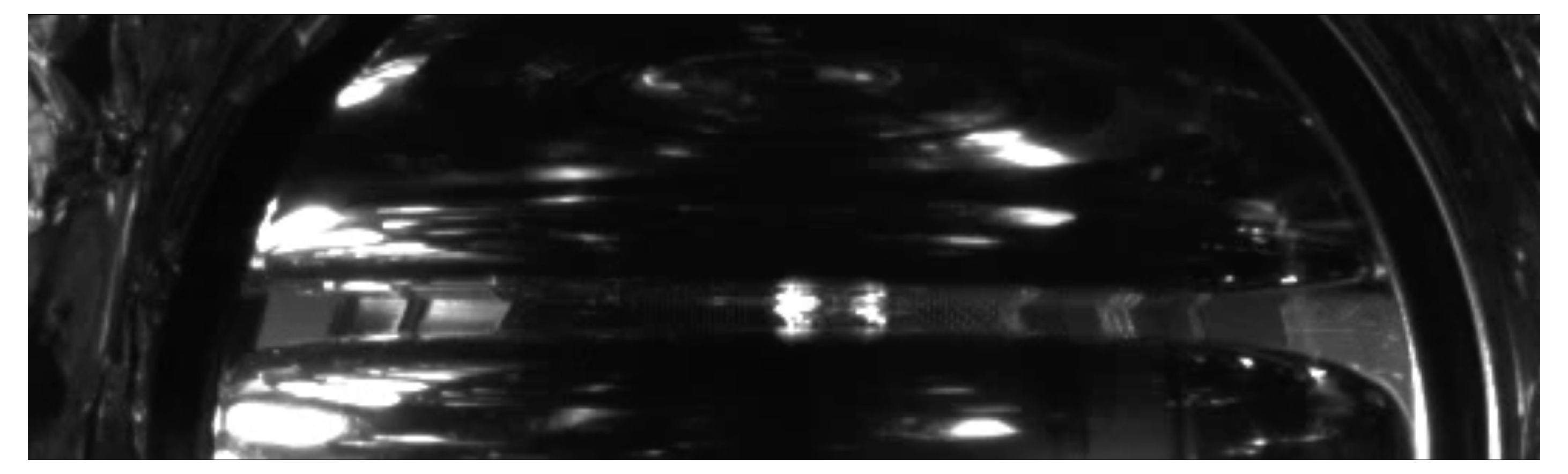}
    
    \includegraphics[trim=0 6cm 0 6cm, clip, width=0.5\columnwidth]{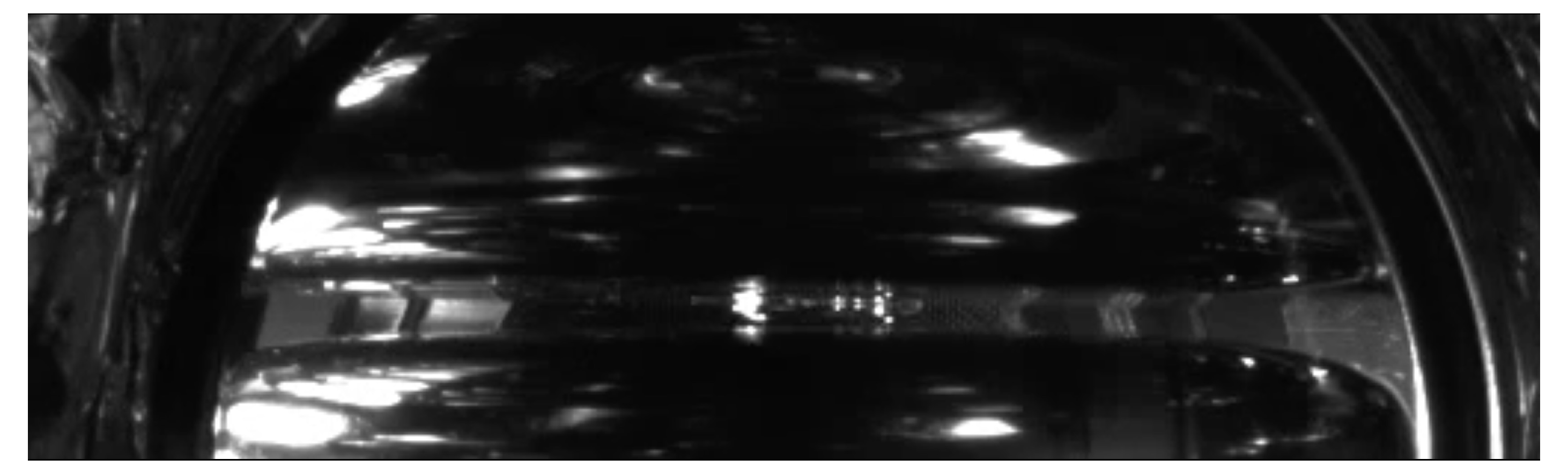}
    \caption{An example of a breakdown containing two apparent discharges. Four frames are shown in sequential order from top to bottom. Two light spots appear in the center of the frame before a big flash of light is emitted and two bubbles emerge from that location immediately after. 
    }
    \label{fig:spark_sequence}
\end{figure}

Breakdowns are associated with a distinct, bright arc and an audible cracking sound (transmitted through the metal) at larger separation distances.
Copious boiling of LXe is usually followed after such event.
More importantly, evidence of some activity developing in the liquid around the breakdown location prior to the main discharge was usually observed. 
These appear as faint, flickering arcs between the electrodes. For instance, \cref{fig:spark_sequence} shows an example of two near-simultaneous breakdowns spawning from two visible arcs and followed by two separate bubbles.

\begin{figure}
    \centering
    \includegraphics[trim=0 6cm 0 6cm, clip, width=0.75\columnwidth]{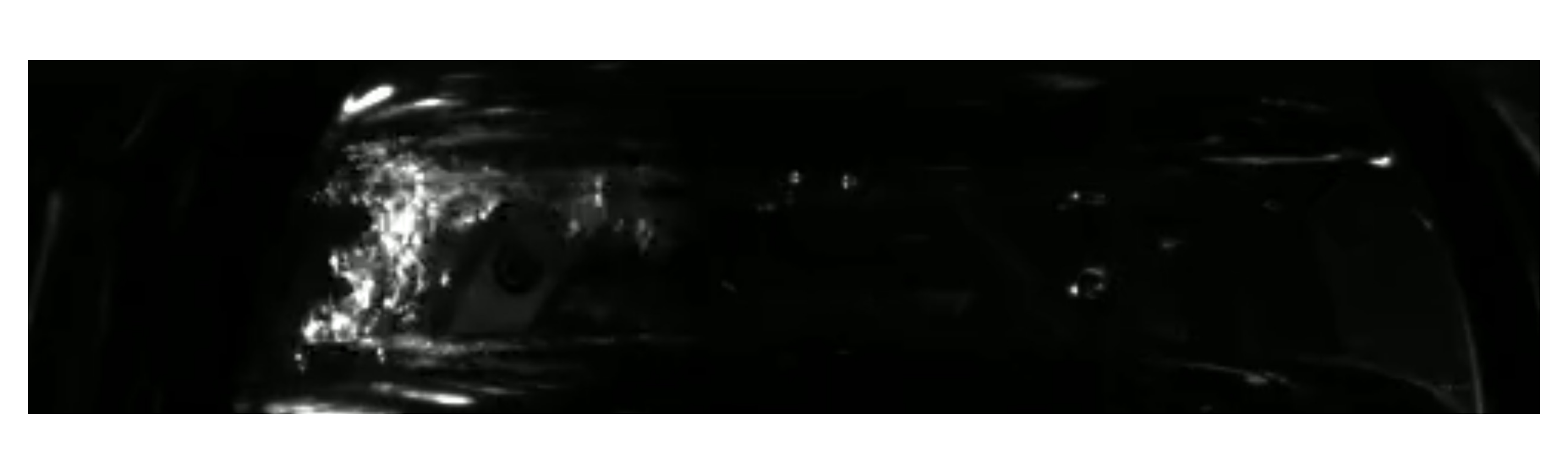}
    \caption{Video of a bubble occurring prior to a breakdown. A small bubble starts forming on the top left corner, travels down to the cathode's surface and a breakdown occurs in that same location (\href{https://drive.google.com/file/d/1EzKIpoTRTheYDjKnccUVnHKk7QmjLaBS/view?usp=sharing}{Multimedia view}).}
    \label{fig:bubble_before_breakdown_video}
\end{figure}

An interesting possibility is that some of these early bright spots correspond to bubbles that nucleated prior to a breakdown.
This would corroborate the hypothesis that bubbles are an essential mechanism for originating dielectric breakdown.
In most breakdowns, bubbles take a second or more to roll off the side of the electrodes after a discharge and they also grow in size over time. 
Therefore, it is unlikely that these bubbles floated in between the electrodes from elsewhere in the apparatus.
One particular good example of a bubble possibly initiating a breakdown is shown in \cref{fig:bubble_before_breakdown_video} (multimedia file). In this short video, one can see a small bubble traveling from the anode to the cathode on the left side of the frame and subsequently a breakdown occurs in that same location.  

Perhaps one of the most intriguing visual observations was that of a \dquotes{shimmering} effect seen during some ramps.
At large separations and during a high voltage ramp, the LXe would begin to ripple, increasing in intensity until a breakdown occurred, and then  settling down to a more placid condition in time for the next ramp.
A close visual analogue for this effect would be the appearance of mixing alcohol and water.
It is worth noticing that this effect was never observed with unbiased electrodes in stable conditions.
We hypothesize that this phenomenon was due to localized heating of the xenon, leading to changes in the index of refraction.
It is possible, as pointed out in Ref.~\cite{atrazhev_mechanisms_2010}, that such heating plays a key role in the development of a discharge. 
If mixing is poor, the liquid may become locally superheated until an instability occurs, leading to spontaneous nucleation of bubbles, which then initiate a dielectric breakdown in the bulk of the detector.

Finally, following several runs of XeBrA, it was noticed that some damage was done to the electrodes in the form of pitting.
The electrodes were examined under a microscope to reveal pits of various sizes on both the cathode and anode.
Subsequent to this, the electrodes were replaced with duplicates and the originals were re-polished.
Following further pitting to the second pair of electrodes similar to the first, separation distances of over 3~mm were avoided to keep the stored energy lower.
This avoided the more significant pitting seen in the initial set of electrodes, although some damage was still done over the course of a run.

\section{\label{sec:Conclusions} Conclusion}

With the XeBrA experiment, which employs a pair of Rogowski electrodes with adjustable separation distance, we confirmed that breakdown field scales inversely with stressed electrode area in LXe for stressed areas up to 33 cm$^{2}$. 
No significant, reproducible scaling with LXe pressure was observed between 1.5 bar and 2.2 bar absolute.
Similarly, no significant dependence was observed with the high voltage ramp speed in the range between \SI{100}{V/s} and \SI{200}{V/s}.
By contrast, a small increase in breakdown field was observed with the passivation of the electrode surfaces.

The location of a breakdown on the electrode surface was reconstructed from high-speed videos.
They show moderate correlations with FN current as obtained by electric field simulations.
Moreover, evidence for bubble nucleation initiating breakdown was observed with low-speed videos.
We hypothesize that the bubble nucleation itself is caused by localized heating due to field emission from field enhancers, such as local asperities. 

Finally, some practical recommendations can be drawn from this work.
First, significant breakdown precursor activity was observed.
Small discharges were detected with a SiPM and a charge-sensitive amplifier connected to the anode.
These discharges slowly increase in intensity in the last 30 seconds before a breakdown and sharply accelerate in the last second. Hence, having the ability to ramp down the voltage on the scale of seconds could be beneficial for avoiding electrical breakdown. Second, a downward trend in current collected at the anode was observed over the course of a run, which motivates conducting an initial conditioning campaign prior to the start of any science data-taking campaign.
Third, the extrapolation from our measured stressed area scaling shows that the breakdown field for the next-generation, LXe TPC experiments, will only be on the order of a few tens of \SI{}{kV/cm}. These results have important implications for the design of the high voltage delivery system of such experiments.

\begin{acknowledgments}
We are grateful to the support received from the technical and engineering staff at LBNL. Also, we thank Takeyasu Ito and Nguyen Phan for helpful correspondence. Funding for this work is supported by the U.S.~Department of Energy, Office of Science, Office of High Energy Physics under Contract Numbers DE-SC0018982 and DE-AC02-05CH11231.
\end{acknowledgments}

\section*{Data Availability Statement}

The data that support the findings of this study are available from the corresponding author upon reasonable request.


\appendix
\onecolumngrid

\section{Plumbing diagram of XeBrA}
\label{appd:xebra_pid}

\begin{sidewaysfigure}
    \centering
    \includegraphics[width=0.9\linewidth]{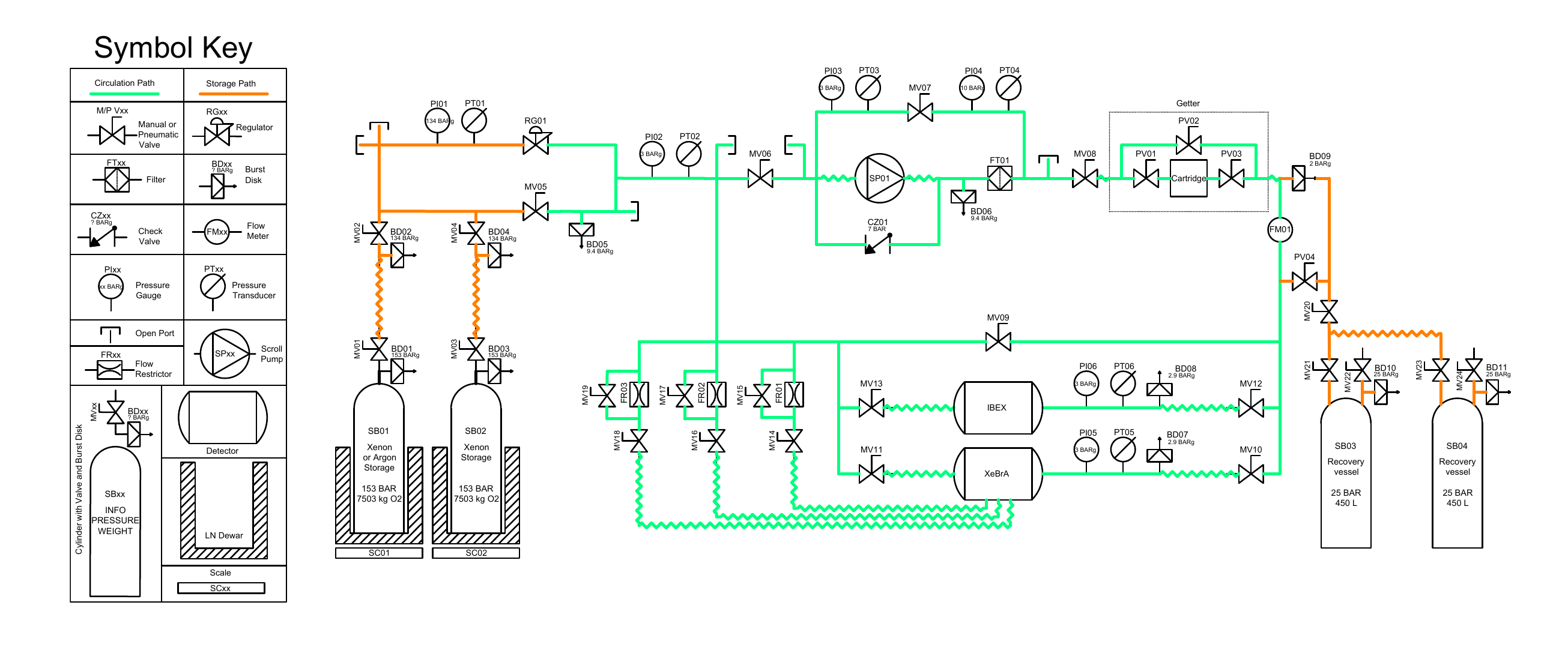}
    \caption{Piping and instrumentation (P\&ID) of the XeBrA experiment.}
    \label{fig:xebra_pid}
\end{sidewaysfigure}

See \cref{fig:xebra_pid}.

\section{Summary of fit results}
\label{appd:run_table}

See \cref{tab:run_fits}

\begin{table}
\begin{tabular}{llllllllllll}
\hline
Dataset & Electrodes &  Tilt &  Gap  &  Ramp Rate  &  Pressure& N & Model&k& $E_0$ &  $E_1$ & $ E_2$ \\
      &  &  [$\circ$] &  [mm]  &  [V/s]  &  [bar]&  &  &&   [kV/cm] &  [kV/cm]  &  [kV/cm]          \\
\hline
       400 &           MP1 &        0.61 &       3.0 &              100 &            1.70 &  19 &           3 &   3.1 $\pm$ 2.3 &    2.8 $\pm$ 1.9 &   28.6 $\pm$ 1.8 &  31.4 $\pm$ 0.2 \\
            401 &           MP1 &        0.61 &       1.0 &              100 &            2.00 &  66 &           3 &   5.6 $\pm$ 3.8 &  37.5 $\pm$ 22.2 &  77.4 $\pm$ 21.3 & 114.9 $\pm$ 1.3 \\
            402 &           MP1 &        0.61 &       2.0 &              100 &            2.00 &  60 &           3 & 16.4 $\pm$ 12.3 &  78.4 $\pm$ 55.4 &   6.3 $\pm$ 54.9 &  84.7 $\pm$ 0.7 \\
            403 &           MP1 &        0.61 &       3.0 &              100 &            2.00 & 145 &           3 &  12.2 $\pm$ 0.0 &   77.1 $\pm$ 0.0 &    0.0 $\pm$ 0.0 &  77.1 $\pm$ 0.0 \\
            404 &           MP1 &        0.61 &       5.0 &              100 &            2.00 &  40 &           1 &  15.4 $\pm$ 1.9 &   71.1 $\pm$ 0.7 &    0.0 $\pm$ 0.0 &  71.1 $\pm$ 0.7 \\
            500 &           MP1 &        0.61 &       1.0 &              100 &            2.01 &  35 &           3 &  44.7 $\pm$ 8.1 &  83.5 $\pm$ 13.0 &   4.9 $\pm$ 13.0 &  88.4 $\pm$ 0.2 \\
            501 &           MP1 &        0.61 &       2.0 &              100 &            2.00 &  81 &           3 &  12.1 $\pm$ 2.8 &   82.4 $\pm$ 1.7 &    0.0 $\pm$ 0.0 &  82.4 $\pm$ 1.7 \\
            504 &           MP1 &        0.61 &       2.0 &              100 &            2.00 &  20 &           1 &   9.8 $\pm$ 1.7 &   61.0 $\pm$ 1.4 &    0.0 $\pm$ 0.0 &  61.0 $\pm$ 1.4 \\
            505 &           MP1 &        0.61 &       3.0 &              100 &            2.00 &  34 &           2 &   1.5 $\pm$ 0.2 &    9.4 $\pm$ 1.3 &   41.5 $\pm$ 0.3 &  50.9 $\pm$ 1.2 \\
            506 &           MP1 &        0.61 &       5.0 &              100 &            1.98 &  97 &           3 &  9.4 $\pm$ 12.6 &  29.0 $\pm$ 46.6 &   3.0 $\pm$ 15.6 & 32.0 $\pm$ 56.0 \\
            601 &           MP1 &        0.28 &       1.0 &              100 &            2.01 &  53 &           3 &   1.9 $\pm$ 0.4 &   20.2 $\pm$ 3.3 &   81.7 $\pm$ 1.6 & 101.8 $\pm$ 1.8 \\
            602 &           MP1 &        0.28 &       2.3 &              100 &            2.01 &  21 &           2 &   1.8 $\pm$ 0.2 &   14.5 $\pm$ 1.9 &   41.0 $\pm$ 0.5 &  55.5 $\pm$ 1.5 \\
            603 &           MP1 &        0.28 &       2.0 &              100 &            2.01 &  38 &           1 &   7.8 $\pm$ 1.0 &   77.2 $\pm$ 1.7 &    0.0 $\pm$ 0.0 &  77.2 $\pm$ 1.7 \\
            701 &           MP1 &        0.28 &       5.0 &              100 &            2.00 &  42 &           3 &  9.8 $\pm$ 16.3 & 53.1 $\pm$ 118.3 & 19.1 $\pm$ 119.9 &  72.1 $\pm$ 4.5 \\
            703 &           MP1 &        0.28 &       1.0 &              100 &            2.00 &  50 &           1 &  11.9 $\pm$ 1.4 &  111.9 $\pm$ 1.3 &    0.0 $\pm$ 0.0 & 111.9 $\pm$ 1.3 \\
            704 &           MP1 &        0.28 &       2.0 &              100 &            2.00 &  37 &           1 &  17.0 $\pm$ 2.2 &   91.5 $\pm$ 0.9 &    0.0 $\pm$ 0.0 &  91.5 $\pm$ 0.9 \\
            705 &           MP1 &        0.28 &       3.0 &              100 &            2.00 &  40 &           3 &  19.2 $\pm$ 2.7 &   85.7 $\pm$ 3.9 &    0.3 $\pm$ 2.9 &  86.0 $\pm$ 1.2 \\
            706 &           MP1 &        0.28 &       2.0 &              100 &            1.70 &  41 &           1 &  17.0 $\pm$ 2.1 &   90.9 $\pm$ 0.8 &    0.0 $\pm$ 0.0 &  90.9 $\pm$ 0.8 \\
            707 &           MP1 &        0.28 &       2.0 &              100 &            2.20 &  44 &           1 &  18.0 $\pm$ 2.1 &   97.9 $\pm$ 0.8 &    0.0 $\pm$ 0.0 &  97.9 $\pm$ 0.8 \\
            803 &           MP2 &        1.04 &       3.0 &              100 &            2.00 &  47 &           1 &  10.2 $\pm$ 1.2 &   79.2 $\pm$ 1.2 &    0.0 $\pm$ 0.0 &  79.2 $\pm$ 1.2 \\
            804 &           MP2 &        1.04 &       2.0 &              100 &            2.00 &  36 &           1 &  18.4 $\pm$ 2.3 &   75.3 $\pm$ 0.7 &    0.0 $\pm$ 0.0 &  75.3 $\pm$ 0.7 \\
            805 &           MP2 &        1.04 &       1.0 &              100 &            2.00 &  36 &           3 &  34.8 $\pm$ 3.9 &   78.4 $\pm$ 3.2 &    0.6 $\pm$ 3.2 &  79.0 $\pm$ 0.3 \\
            806 &           MP2 &        1.04 &       3.0 &              100 &            2.00 &  18 &           1 &  12.0 $\pm$ 2.0 &   67.2 $\pm$ 1.4 &    0.0 $\pm$ 0.0 &  67.2 $\pm$ 1.4 \\
            807 &           MP2 &        1.04 &       2.0 &              100 &            2.00 &  18 &           3 & 25.4 $\pm$ 22.9 &   80.3 $\pm$ 1.9 &    0.0 $\pm$ 1.7 &  80.3 $\pm$ 1.8 \\
            808 &           MP2 &        1.04 &       1.0 &              100 &            2.00 &  33 &           2 &   1.6 $\pm$ 0.1 &    3.6 $\pm$ 0.3 &   69.2 $\pm$ 0.1 &  72.8 $\pm$ 0.3 \\
            809 &           MP2 &        1.04 &       2.0 &              100 &            1.50 &  42 &           3 &   3.6 $\pm$ 0.7 &    7.8 $\pm$ 1.4 &   72.5 $\pm$ 1.0 &  80.4 $\pm$ 0.5 \\
            810 &           MP2 &        1.04 &       2.0 &              100 &            1.60 &  37 &           3 & 51.1 $\pm$ 23.3 &  66.7 $\pm$ 27.6 &  10.9 $\pm$ 27.5 &  77.6 $\pm$ 0.3 \\
            811 &           MP2 &        1.04 &       2.0 &              100 &            1.70 &  34 &           3 &   2.6 $\pm$ 0.6 &    6.6 $\pm$ 0.9 &   71.9 $\pm$ 0.5 &  78.5 $\pm$ 0.6 \\
            812 &           MP2 &        1.04 &       2.0 &              100 &            1.80 &  38 &           3 &  19.5 $\pm$ 9.6 &  59.9 $\pm$ 26.6 &  17.9 $\pm$ 26.4 &  77.8 $\pm$ 0.9 \\
            813 &           MP2 &        1.04 &       5.0 &              100 &            2.00 &  39 &           1 &  13.3 $\pm$ 1.6 &   83.6 $\pm$ 1.0 &    0.0 $\pm$ 0.0 &  83.6 $\pm$ 1.0 \\
            900 &           MP1 &        1.26 &       3.0 &              100 &            2.00 &  29 &           1 &  13.2 $\pm$ 2.0 &   86.5 $\pm$ 1.2 &    0.0 $\pm$ 0.0 &  86.5 $\pm$ 1.2 \\
            901 &           MP1 &        1.26 &       3.0 &              100 &            2.00 &  47 &           1 &  13.1 $\pm$ 1.4 &   96.5 $\pm$ 1.1 &    0.0 $\pm$ 0.0 &  96.5 $\pm$ 1.1 \\
            902 &           MP1 &        1.26 &       1.0 &              100 &            2.00 &  67 &           3 &  10.6 $\pm$ 4.0 &  134.0 $\pm$ 4.5 &    0.0 $\pm$ 0.0 & 134.0 $\pm$ 4.5 \\
            903 &           MP1 &        1.26 &       2.0 &              100 &            2.00 &  81 &           1 &  10.1 $\pm$ 0.9 &   96.2 $\pm$ 1.1 &    0.0 $\pm$ 0.0 &  96.2 $\pm$ 1.1 \\
            904 &           MP1 &        1.26 &       2.0 &              150 &            2.00 &  33 &           1 &  16.0 $\pm$ 2.0 &  106.2 $\pm$ 1.2 &    0.0 $\pm$ 0.0 & 106.2 $\pm$ 1.2 \\
            905 &           MP1 &        1.26 &       2.0 &              200 &            2.00 &  35 &           1 &  10.2 $\pm$ 1.3 &   98.1 $\pm$ 1.7 &    0.0 $\pm$ 0.0 &  98.1 $\pm$ 1.7 \\
            906 &           MP1 &        1.26 &       2.0 &              100 &            1.70 &  34 &           1 &  13.7 $\pm$ 1.7 &  107.9 $\pm$ 1.4 &    0.0 $\pm$ 0.0 & 107.9 $\pm$ 1.4 \\
            907 &           MP1 &        1.26 &       2.0 &              100 &            1.50 &  40 &           1 &  13.8 $\pm$ 1.6 &  104.0 $\pm$ 1.2 &    0.0 $\pm$ 0.0 & 104.0 $\pm$ 1.2 \\
           1000 &         Pass. &        0.51 &       3.0 &              100 &            2.00 &  19 &           1 &  14.1 $\pm$ 2.4 &   78.9 $\pm$ 1.3 &    0.0 $\pm$ 0.0 &  78.9 $\pm$ 1.3 \\
           1001 &         Pass. &        0.51 &       3.0 &              100 &            2.00 &  40 &           2 &   2.2 $\pm$ 0.4 &   28.6 $\pm$ 3.7 &   46.2 $\pm$ 1.9 &  74.8 $\pm$ 2.3 \\
           1002 &         Pass. &        0.51 &       1.0 &              100 &            2.00 &  71 &           3 &  15.4 $\pm$ 2.5 &  148.6 $\pm$ 2.3 &    0.0 $\pm$ 0.1 & 148.6 $\pm$ 2.3 \\
           1003 &         Pass. &        0.51 &       2.0 &              100 &            2.00 &  78 &           3 &   8.6 $\pm$ 3.8 &  103.6 $\pm$ 6.1 &    0.0 $\pm$ 0.0 & 103.6 $\pm$ 6.1 \\
           1004 &         Pass. &        0.51 &       2.0 &              150 &            2.00 &  48 &           3 &  13.8 $\pm$ 11.3 & 113.5 $\pm$ 5.1 &    0.0 $\pm$ 0.0 & 113.5 $\pm$ 5.1 \\
           1005 &         Pass. &        0.51 &       2.0 &              200 &            2.00 &  38 &           1 &  12.3 $\pm$ 1.6 &  117.5 $\pm$ 1.6 &    0.0 $\pm$ 0.0 & 117.5 $\pm$ 1.6 \\
           1006 &         Pass. &        0.51 &       2.0 &              100 &            1.60 &  42 &           1 &  12.8 $\pm$ 1.5 &  102.0 $\pm$ 1.3 &    0.0 $\pm$ 0.0 & 102.0 $\pm$ 1.3 \\
           1007 &         Pass. &        0.51 &       2.0 &              100 &            1.50 &  40 &           1 &  15.2 $\pm$ 1.8 &  117.0 $\pm$ 1.2 &    0.0 $\pm$ 0.0 & 117.0 $\pm$ 1.2 \\
           1008 &         Pass. &        0.51 &       2.0 &              100 &            1.70 &  42 &           1 &  13.9 $\pm$ 1.7 &  122.7 $\pm$ 1.4 &    0.0 $\pm$ 0.0 & 122.7 $\pm$ 1.4 \\
\hline
\end{tabular}
\caption{The results of the Weibull fitting procedure on a per-dataset basis. The last two digits of the first column indicates the dataset number and the number in front represents the run number. Three sets of electrodes were employed: two mechanically polished (MP1 and MP2) and one with passivated surfaces (Pass.). The cathode tilt angle, electrode gap separation, ramp rate, and pressure for each dataset are also indicated. Column \dquotes{N} shows the number of breakdown events contained in each dataset after applying data quality cuts and column \dquotes{Model} indicates the model that was selected to fit the breakdown field data (following the procedure described in \cref{sec:Stressed Area}).  
Datasets with less than 15 breakdown events were discarded, amounting to 10 in total.
7 additional datasets were discarded because of unfavorable conditions were found with the experimental setup after the start of data acquisition.
The best-fit for the Weibull model parameters (defined in \cref{sec:hazard_functions}) are shown in the last 4 columns.}
\label{tab:run_fits}
\end{table}

\newpage
\bibliographystyle{apsrev4-1}
\bibliography{sources}

\end{document}